\documentclass[abstract,headings=normal]{scrartcl}

\pdfoutput=1

\usepackage[automark]{scrpage2}
\usepackage[english]{babel}
\usepackage[T1]{fontenc}
\usepackage[applemac]{inputenc}
\usepackage{graphicx}
\usepackage{tikz}
\usepackage{amsmath}
\usepackage{amssymb}
\usepackage{amsthm}
\usepackage{subfig}
\usepackage{hyperref}

\newcommand{\C}{\mathbb{C}}
\newcommand{\Z}{\mathbb{Z}}

\newcommand{\T}{\mathcal{T}}

\newcommand{\Hvier}{H\textsuperscript{4}}
\newcommand{\Hsechs}{H\textsuperscript{6}}

\newtheorem{theo}{Theorem}[section]
\newtheorem{lemma}[theo]{Lemma}
\newtheorem{cor}[theo]{Corollary}
\theoremstyle{definition}
\newtheorem{defi}[theo]{Definition}
\newtheorem{const}[theo]{Construction}
\theoremstyle{remark}
\newtheorem*{rem}{Remark}

\graphicspath{{./graphics/}}
\DeclareGraphicsExtensions{.pdf}

\begin{document}

\title{On Bianchi permutability of B\"acklund~transformations for asymmetric~quad-equations}

\author{Raphael Boll\footnote{Institut f\"ur Mathematik, MA 7-2, Technische Universit\"at Berlin, Str.~des~17.~Juni~136, 10623 Berlin, Germany; e-mail: boll@math.tu-berlin.de}}

\maketitle
\begin{abstract} We prove the Bianchi permutability (existence of superposition principle) of B\"acklund transformations for asymmetric quad-equations. Such equations and there B\"acklund transformations form 3D consistent systems of a priori different equations. We perform this proof by using 4D~consistent systems of quad-equations, the structural insights through biquadratics patterns and the consideration of super-consistent eight-tuples of quad-equations on decorated cubes.\par
\vspace{0.5cm}
\end{abstract}

\section{Introduction}
One of the definitions of integrability of discrete equations, which becomes increasingly popular in the recent years, is based on the notion of multidimensional consistency. For two-dimensional systems, this notion was clearly formulated first in \cite{NW}, and it was pushed forward as a definition of integrability in \cite{quadgraphs,Nijhoff}. The outstanding importance of 3D consistency in the theory of discrete integrable systems became evident no later than with the appearance of the well-known ABS-classification of integrable equations in \cite{ABS1}. This classification deals with quad-equations which can be set on the faces of an elementary cube in such a manner that all faces carry similar equations differing only by the parameter values assigned to the edges of a cube. Moreover, each single equation admits a $D_4$ symmetry group. Thus, ABS-equations from \cite{ABS1} can be extended in a straightforward manner to the whole of $\Z^{m}$.  In \cite{ABS2}, a relaxed definition of 3D consistency was introduced: the faces of an elementary cube are allowed to carry a priori different quad-equations. The classification performed in \cite{ABS2} is restricted to so-called equations of type~Q, i.e., those equations whose biquadratics are all non-degenerate (a precise definition will be recalled in Section \ref{faces}). In \cite{Atk1} and \cite{Todapaper}, numerous asymmetric systems of quad-equations have been studied, which also include equations of type H (with degenerate biquadratics). A classification of such systems has been given in  \cite{classification}.  This classification covers the majority of systems from \cite{Atk1} and all systems from \cite{Todapaper}, as well as equations from \cite{Hietarinta}, \cite{LY} and \cite{HV}. Moreover, it contains also a number of novel systems. The results of this local classification lead to integrable lattice systems via a procedure of reflecting the cubes in a suitable way (see \cite{classification} for details).\par
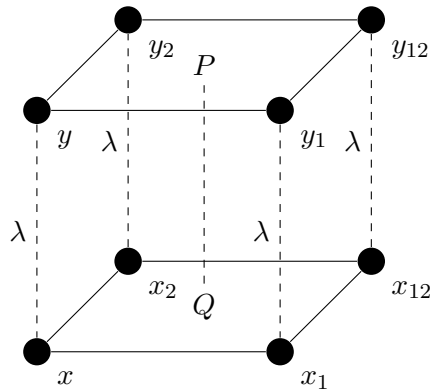
\begin{figure}[htbp]
   \centering
   \begin{tikzpicture}[auto,scale=0.8]
      \node (x) at (0,0) [circle,fill,label=-45:$x$] {};
      \node (x1) at (4,0) [circle,fill,label=-45:$x_{1}$] {};
      \node (x2) at (1.5,1.5) [circle,fill,label=-45:$x_{2}$] {};
      \node (x3) at (0,4) [circle,fill,label=-45:$y$] {};
      \node (x12) at (5.5,1.5) [circle,fill,label=-45:$x_{12}$] {};
      \node (x13) at (4,4) [circle,fill,label=-45:$y_{1}$] {};
      \node (x23) at (1.5,5.5) [circle,fill,label=-45:$y_{2}$] {};
      \node (x123) at (5.5,5.5) [circle,fill,label=-45:$y_{12}$] {};
      \node (A) at (2.75,0.75) {$Q$};
      \node (Aq) at (2.75,4.75) {$P$};
      \draw (x) to (x1);
      \draw (x1) to (x12);
      \draw [dashed] (x12) to node {$\lambda$} (x123);
      \draw (x123) to (x23);
      \draw (x23) to (x3);
      \draw [dashed] (x3) to node [swap] {$\lambda$} (x);
      \draw (x3) to (x13);
      \draw [dashed] (x13) to node [swap] {$\lambda$} (x1);
      \draw (x13) to (x123);
      \draw (x) to (x2);
      \draw (x2) to (x12);
      \draw [dashed] (x2) to node {$\lambda$} (x23);
      \draw [dashed] (A) to (Aq);
   \end{tikzpicture}
   \caption{Elementary 3D cube of a B\"acklund transformation $x\overset{\lambda}{\mapsto}y$, a B\"acklund cube}
   \label{fig:H1Baecklund}
\end{figure}

As already mentioned in \cite{BS1,classification}, one of the reasons for using 3D consistency as a definition for integrability is that one can derive B\"acklund transformations and zero-curvature representations from 3D consistency. In the present paper we will have a closer look to B\"acklund transformations. We have to properly generalize the definition of B\"acklund transformations (see Definition~\ref{def:Baecklund} on page~\pageref{def:Baecklund}), because the quad-equations of the classification of \cite{classification} have less symmetry than the ones of \cite{ABS1}. In fact, such a B\"acklund transformation $x\overset{\lambda}{\mapsto} y$ between solutions $x$ and $y$ can be described by a set of six quad-equations which is called \emph{B\"acklund cube} and can be assigned to the faces of a combinatorial cube such that 3D consistency is satisfied. Here, the equation on the bottom face is the original equation $Q\left(x,x_{1},x_{2},x_{12}\right)=0$, the equation on the top face is the B\"acklund transform $P\left(y,y_{1},y_{2},y_{12}\right)=0$ and the four equations on the vertical faces describe the B\"acklund transformation itself (see Figure~\ref{fig:H1Baecklund}). The so-called B\"acklund parameter is denoted by $\lambda$.\par
\begin{figure}[htbp]
   \centering
   \begin{tikzpicture}[auto,>=stealth]
      \node (v) at (0,0) {$x$};
      \node (v1) at (2,-1) {$y^{\left(1\right)}$};
      \node (v2) at (2,1) {$y^{\left(2\right)}$};
      \node (v12) at (4,0) {$X$};
      \draw [ultra thick,->] (v) to node [swap] {$\lambda_{1}$} (v1);
      \draw [ultra thick,->] (v) to node {$\lambda_{2}$} (v2);
      \draw [ultra thick,->] (v2) to node {$\lambda_{1}$} (v12);
      \draw [ultra thick,->] (v1) to node [swap] {$\lambda_{2}$} (v12);
   \end{tikzpicture}
   \caption{Superposition principle of four solutions $x,y^{\left(1\right)},y^{\left(2\right)},X:\Z^{2}\to\C$, a Bianchi diagram}
   \label{fig:H1Bianchi}
\end{figure}
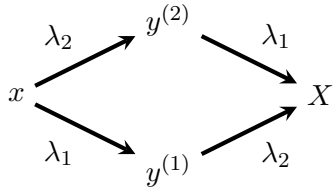

A very important feature of B\"acklund transformations is the so-called Bianchi permutability \cite{bianchi} or the existence of a superposition principle. We will show that, given two B\"acklund transformations $x\overset{\lambda_{1}}\mapsto y^{\left(1\right)}$ and $x\overset{\lambda_{2}}{\mapsto} y^{\left(2\right)}$, there exists a unique, in some sense, function $X$ which is a B\"acklund transform of $y^{\left(1\right)}$ and $y^{\left(2\right)}$ simultaneously (see Figure~\ref{fig:H1Bianchi}).\par
\begin{figure}[htbp]
   \centering
   \begin{tikzpicture}[auto,scale=0.8]
      \node (x) at (0,0) [circle,fill,label=-45:$x$] {};
      \node (x1) at (4,0) [circle,fill,label=-45:$x_{1}$] {};
      \node (x2) at (1.5,1.5) [circle,fill,label=-45:$x_{2}$] {};
      \node (x12) at (5.5,1.5) [circle,fill,label=-45:$x_{12}$] {};
      \node (x3) at (-3.5,4) [circle,fill,label=180:$y^{\left(1\right)}$] {};
      \node (x13) at (0.5,4) [circle,fill,label=0:$y^{\left(1\right)}_{1}$] {};
      \node (x23) at (-2,5.5) [circle,fill,label=180:$y^{\left(1\right)}_{2}$] {};
      \node (x123) at (2,5.5) [circle,fill,label=0:$y^{\left(1\right)}_{12}$] {};
      \node (x4) at (3,4) [circle,fill,label=180:$y^{\left(2\right)}$] {};
      \node (x14) at (7,4) [circle,fill,label=0:$y^{\left(2\right)}_{1}$] {};
      \node (x24) at (4.5,5.5) [circle,fill,label=180:$y^{\left(2\right)}_{2}$] {};
      \node (x124) at (8.5,5.5) [circle,fill,label=0:$y^{\left(2\right)}_{12}$] {};
      \node (x34) at (-0.5,8) [circle,fill,label=135:$X$] {};
      \node (x134) at (3.5,8) [circle,fill,label=135:$X_{1}$] {};
      \node (x234) at (1,9.5) [circle,fill,label=135:$X_{2}$] {};
      \node (x1234) at (5,9.5) [circle,fill,label=135:$X_{12}$] {};
      \draw (x) to (x1);
      \draw (x1) to (x12);
      \draw (x) to (x2);
      \draw (x2) to (x12);
      \draw (x123) to (x23);
      \draw (x23) to (x3);
      \draw (x3) to (x13);
      \draw (x13) to (x123);
      \draw (x124) to (x24);
      \draw (x24) to (x4);
      \draw (x4) to (x14);
      \draw (x14) to (x124);
      \draw (x34) to (x134);
      \draw (x134) to (x1234);
      \draw (x34) to (x234);
      \draw (x234) to (x1234);
      \draw [dashed] (x12) to (x123);
      \draw [dashed] (x3) to (x);
      \draw [dashed] (x13) to (x1);
      \draw [dashed] (x2) to (x23);
      \draw [dashed] (x12) to (x124);
      \draw [dashed] (x4) to (x);
      \draw [dashed] (x14) to (x1);
      \draw [dashed] (x2) to (x24);
      \draw [dashed] (x1234) to (x123);
      \draw [dashed] (x3) to (x34);
      \draw [dashed] (x13) to (x134);
      \draw [dashed] (x234) to (x23);
      \draw [dashed] (x1234) to (x124);
      \draw [dashed] (x4) to (x34);
      \draw [dashed] (x14) to (x134);
      \draw [dashed] (x234) to (x24);
   \end{tikzpicture}
   \caption{Extension of two B\"acklund cubes which share the equation for $\left(x,x_{1},x_{2},x_{12}\right)$ and correspond to the elementary cubes of B\"acklund transformations $x\overset{\lambda_{1}}{\mapsto}y^{\left(1\right)}$ and $x\overset{\lambda_{2}}{\mapsto}y^{\left(2\right)}$. The result is a 4D consistent 24-tuple of quad-equations each of the 24 2D faces carrying a quad-equation.}
   \label{fig:H1Super}
\end{figure}
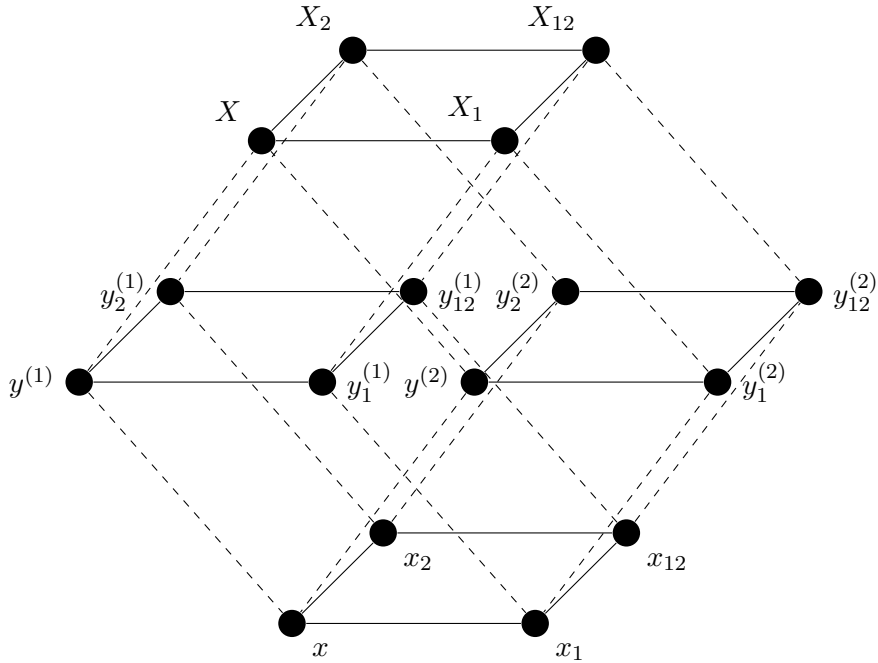

In other words, we will show the following theorem which will be the first main result of this paper:
\begin{theo}\label{th:Bianchi}
Given two B\"acklund cubes sharing one equation satisfying the tetrahedron property and containing no equation of type~\Hsechs (see beginning of Section~\ref{classi} for more explanation of these conditions). Then there exists an extension to a 4D consistent 24-tuple of quad-equations. This extension is unique up to M\"obius transformations of fields not belonging to the original B\"acklund cubes (the fields $X$, $X_{1}$, $X_{2}$ and $X_{12}$ in Figure~\ref{fig:H1Super}).
\end{theo}
As an example of an integrable asymmetric quad-equation we consider the equation
\begin{equation}\label{eq:H1}
\left(x-x_{1}\right)\left(x_{2}-x_{12}\right)-\alpha\left(1+\epsilon^{2}x_{2}x_{12}\right)=0
\end{equation}
which is the equation $H_{1}^{\epsilon}$ in the trapezoidal version from the classification in \cite{classification}. In this case, asymmetry means that the above equation do only have the symmetry $x\leftrightarrow x_{1}$, $x_{2}\leftrightarrow x_{12}$ and not, e.g., $x\leftrightarrow x_{2}$, $x_{1}\leftrightarrow x_{12}$.\par
In what sense can this equation be considered as integrable? Upon applying the procedure of reflecting the squares it leads to well-defined system on $\Z^{2}$ which, as proven in \cite{classification}, possesses a zero-curvature representation. In the present paper we show that it also possesses
\begin{itemize}
\item elementary non-auto-B\"acklund transformations
\item and two-stage auto-B\"acklund transformations.
\end{itemize}\par
There turn out to exist two essentially different B\"acklund transformations. The first one, $BT_{1}:x\overset{\lambda_{1}}{\mapsto} y^{\left(1\right)}$ can be described by the following equations on the vertical faces:
\begin{equation}\label{eq:H1Baeck1}
\begin{gathered}
\left(x-y^{\left(1\right)}\right)\left(x_{2}-y_{2}^{\left(1\right)}\right)-\lambda_{1}\left(1+\epsilon^{2}x_{2}y_{2}^{\left(1\right)}\right)=0,\\
\lambda_{1}\left(x-x_{1}\right)\left(y^{\left(1\right)}-y_{1}^{\left(1\right)}\right)-\alpha\left(x-y^{\left(1\right)}\right)\left(x_{1}-y_{1}^{\left(1\right)}\right)+\epsilon^{2}\lambda_{1}\alpha\left(\alpha-\lambda_{1}\right)=0,\\
\left(x_{1}-y_{1}^{\left(1\right)}\right)\left(x_{12}-y_{12}^{\left(1\right)}\right)-\lambda_{1}\left(1+\epsilon^{2}x_{12} y_{12}^{\left(1\right)}\right)=0,\\
\lambda_{1}\left(x_{2}-x_{12}\right)\left(y_{2}^{\left(1\right)}-y_{12}^{\left(1\right)}\right)-\alpha\left(x_{2}-y_{2}^{\left(1\right)}\right)\left(x_{12}-y_{12}^{\left(1\right)}\right)=0.
\end{gathered}
\end{equation}
The B\"acklund transform $y^{\left(1\right)}$ satisfies the equation
\begin{equation}\label{eq:H1Baeckt1}
\left(y^{\left(1\right)}-y_{1}^{\left(1\right)}\right)\left(y_{2}^{\left(1\right)}-y_{12}^{\left(1\right)}\right)-\alpha\left(1+\epsilon^{2}y_{2}^{\left(1\right)}y_{12}^{\left(1\right)}\right)=0,
\end{equation}
so that $BT_{1}$ is an auto-B\"acklund transformation. The second B\"acklund transformation, $BT_{2}:x\overset{\lambda_{2}}{\mapsto} y^{\left(2\right)}$ can be described by the following equations on the vertical faces:
\begin{equation}\label{eq:H1Baeck2}
\begin{aligned}
\left(x-y_{2}^{\left(2\right)}\right)\left(x_{2}-y^{\left(2\right)}\right)-\lambda_{2}\left(1+\epsilon^{2}x_{2}y^{\left(2\right)}\right)&=0,\\
\left(x-x_{1}\right)\left(y^{\left(2\right)}-y_{1}^{\left(2\right)}\right)-\alpha\left(1+\epsilon^{2}y^{\left(2\right)}y_{1}^{\left(2\right)}\right)&=0,\\
\left(x_{1}-y_{12}^{\left(2\right)}\right)\left(x_{12}-y_{1}^{\left(2\right)}\right)-\lambda_{2}\left(1+\epsilon^{2}x_{12}y_{1}^{\left(2\right)}\right)&=0,\\
\left(x_{2}-x_{12}\right)\left(y_{2}^{\left(2\right)}-y_{12}^{\left(2\right)}\right)-\alpha\left(1+\epsilon^{2}x_{2}x_{12}\right)&=0.
\end{aligned}
\end{equation}
The B\"acklund transform $y^{\left(2\right)}$ satisfies the equation
\begin{equation}\label{eq:H1Baeckt2}
\left(y^{\left(2\right)}-y_{1}^{\left(2\right)}\right)\left(y_{2}^{\left(2\right)}-y_{12}^{\left(2\right)}\right)-\alpha\left(1+\epsilon^{2}y^{\left(2\right)}y_{1}^{\left(2\right)}\right)=0,
\end{equation}
so that $BT_{2}$ is a non-auto-B\"acklund transformation.\par
We will show that in this situation there is still a superposition principle which can be illustrated by Figure~\ref{fig:H1Bianchi} which says that there exists precisely one function $X$ which is a B\"acklund transform of $y^{\left(1\right)}$ and $y^{\left(2\right)}$ simultaneously and, moreover, an auto-B\"acklund transformation for $y^{\left(2\right)}$, that is, satisfies
\begin{equation}\label{eq:H1Bianchi}
\left(X-X_{1}\right)\left(X_{2}-X_{12}\right)-\alpha\left(1+\epsilon^{2}XX_{1}\right)=0.
\end{equation}
We will later see that the relations (quad-equations) between $x$, $y^{\left(1\right)}$, $y^{\left(2\right)}$ and $X$ depends on the lattice points, see \eqref{eq:H1super}.\par
The second main result shows how to produce auto-B\"acklund transformations for asymmetric quad-equations from (a priori) non-auto-B\"acklund transformations (see Construction~\ref{th:autoBaeck} on page~\pageref{th:autoBaeck}). We will show that a two-stage application of an arbitrary B\"acklund transformation (accompanied by a certain reflection procedure illustrated in Figure~\ref{fig:cube4} on page \pageref{fig:cube4}) leads to an auto-B\"acklund transformation. In Theorem~\ref{th:Bianchi2} on page~\pageref{th:Bianchi2} we show two such auto-B\"acklund transformations $x^{\left(1\right)}$ and $x^{\left(2\right)}$ possess a superposition principle expressed by $x^{\left(12\right)}=x^{\left(21\right)}$.\par
The outline of the present paper is as follows: In Sections~\ref{faces} and \ref{cubes} we will introduce notations and recapitulate the classification results from \cite{classification} which are relevant for the present paper. Moreover, in Section~\ref{cubes} we will introduce the generalized definition of a B\"acklund transformation and formulate a procedure how to derive auto-B\"acklund transformations from these transformations. One of the main ideas of this paper is the usage of a new object (see \cite{lagrangian}) which we call a \emph{super-consistent eight-tuple} on a \emph{decorated 3D cube}. Then in Section~\ref{consi} we will introduce the concept of 4D consistency and we will give a proof that this is a direct consequence of 3D consistency. Moreover, we will describe the embedding of quad-equations in four-dimensional lattices. In Section~\ref{classi} we will describe and prove the Bianchi permutability of B\"acklund transformations and, finally, we will give a proof of the Bianchi permutability of auto-B\"acklund transformations in Section~\ref{auto}.

\section{Quad-equations} \label{faces}

\begin{figure}[htbp]
   \centering
   \subfloat[Rhombic version of a quad-equation of type \Hvier, non-degenerate biquadratics on diagonals]{\label{fig:2a}
   \begin{tikzpicture}[auto]
      \node (x1) at (0,0) [circle,fill,label=-135:$x_{1}$] {};
      \node (x4) at (0,2.5) [circle,draw,ultra thick,label=135:$x_{4}$] {};
      \node (x2) at (2.5,0) [circle,draw,ultra thick,label=-45:$x_{2}$] {};
      \node (x3) at (2.5,2.5) [circle,fill,label=45:$x_{3}$] {};
      \draw [thin] (x2) to node {$\alpha_{1}$} (x1);
      \draw [thin] (x4) to node {$\alpha_{1}$} (x3);
      \draw [thin] (x3) to node {$\alpha_{2}$} (x2);
      \draw [thin] (x1) to node {$\alpha_{2}$} (x4);
      \draw [ultra thick] (x2) to (x4);
      \draw [ultra thick] (x1) to (x3);
   \end{tikzpicture}
   }\qquad
   \subfloat[Trapezoidal version of a quad-equation of type \Hvier, non-degenerate biquadratics on one pair of opposite edges]{\label{fig:2b}
   \begin{tikzpicture}[auto]
      \node (x1) at (0,0) [circle,fill,label=-135:$x_{1}$] {};
      \node (x4) at (0,2.5) [circle,draw,ultra thick,label=135:$x_{4}$] {};
      \node (x2) at (2.5,0) [circle,fill,label=-45:$x_{2}$] {};
      \node (x3) at (2.5,2.5) [circle,draw,ultra thick,label=45:$x_{3}$] {};
      \draw [ultra thick] (x2) to node {$\alpha_{1}$} (x1);
      \draw [ultra thick] (x4) to node {$\alpha_{1}$} (x3);
      \draw [thin] (x3) to node {$\alpha_{2}$} (x2);
      \draw [thin] (x1) to node {$\alpha_{2}$} (x4);
      \draw [thin] (x2) to (x4);
      \draw [thin] (x1) to (x3);
   \end{tikzpicture}
   }
   \caption{Biquadratics patterns of type~\Hvier\ equations; non-degenerate biquadratics are indicated by thick lines}
\label{fig:2}
\end{figure}
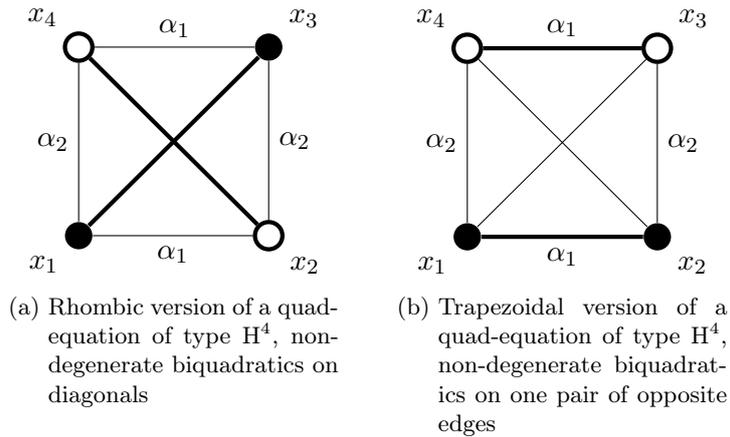

We start with introducing relevant objects and notations. A \emph{quad-equation} is a relation of the type
\[
Q\left(x_{1},x_{2},x_{3},x_{4}\right)=0,\]
where $Q\in\C\left[x_{1},x_{2},x_{3},x_{4}\right]$ is an irreducible multi-affine polynomial. It is convenient to visualize a quad-equation by an elementary square whose vertices carry the four fields $x_1,x_2,x_3,x_4$, cf.~Figure~\ref{fig:2}. For quad-equations without pre-supposed symmetries, the natural classification problem (see \cite{ABS2,classification}) is posed modulo the action of four independent M\"obius transformations of all four variables:
\[
Q(x_1,x_2,x_3,x_4)\; \rightsquigarrow\;
\prod_{k=1}^4(c_kx_k+d_k)\cdot Q\left(\frac{a_1x_1+b_1}{c_1x_1+d_1},\frac{a_2x_2+b_2}{c_2x_2+d_2},
\frac{a_3x_3+b_3}{c_3x_3+d_3},\frac{a_4x_4+b_4}{c_4x_4+d_4}\right).
\]
In the simplest situation, a quad-equation is thought of as an elementary building block of a discrete system $Q(x_{m,n},x_{m+1,n},x_{m+1,n+1},x_{m,n+1})=0$ for a function $x:\Z^2\to\C$. However, if $Q$ is an asymmetric equation, such a simple-minded extension to $\Z^{2}$ does not have to be integrable in any sense, in particular, it does not have to be extendable in the third dimension in a consistent way. Therefore, we will consider a more tricky way of composing discrete systems on $\Z^2$ from different quad-equations:
\begin{const}[Reflecting one quad-equation]\label{const1}
For a quad-equation
\[
Q\left(x,x_{1},x_{2},x_{12};\alpha_{1},\alpha_{2}\right)=0
\]
the three other quad-equations of an elementary cell are given by $\left|Q\right.=0$, $\underline{Q}=0$ and $\left|\underline{Q}\right.=0$ with
\begin{align*}
\left|Q\right.\left(x,x_{1},x_{2},x_{12};\alpha_{1},\alpha_{2}\right)&:=Q\left(x_{1},x,x_{12},x_{2};\alpha_{1},\alpha_{2}\right),\\
\underline{Q}\left(x,x_{1},x_{2},x_{12};\alpha_{1},\alpha_{2}\right)&:=Q\left(x_{2},x_{12},x,x_{1};\alpha_{1},\alpha_{2}\right)\quad \text{and}\\
\left|\underline{Q}\right.\left(x,x_{1},x_{2},x_{12};\alpha_{1},\alpha_{2}\right)&:=Q\left(x_{12},x_{2},x_{1},x;\alpha_{1},\alpha_{2}\right).
\end{align*}
The arguments of the polynomials $\left|Q\right.$, $\underline{Q}$ and $\left|\underline{Q}\right.$ on the neighboring squares can be seen in Figure~\ref{fig:6b}. Parameters $\alpha_{i}$ on edges are distributed so that they are constant along strips of quadrilaterals (i.e., are equal on opposite edges of every elementary quadrilateral). The whole lattice $\Z^{2}$ can be filled by translating this elementary $2\times2$-cell (see Figure~\ref{fig:6a}).
\end{const}
\begin{rem} In the case of equations which do not have parameters which can be assigned to the edges, we have to omit the parameters in the above considerations. By the way, in equation \eqref{eq:H1} -- the example in the introduction -- the parameter $\alpha_{1}$ in the present notation is denoted by $\alpha$, because there is no parameter in this equation which can be assigned to the edge $\left(x,x_{2}\right)$.
\end{rem}
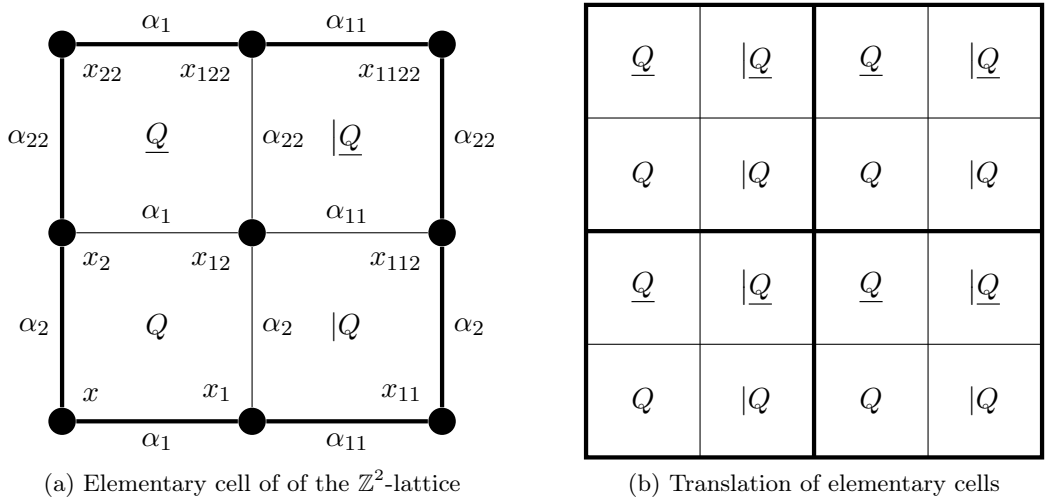
\begin{figure}[htbp]
   \centering
   \subfloat[Elementary cell of of the $\Z^{2}$-lattice]{\label{fig:6b}
   \begin{tikzpicture}[auto]
      \node (Q) at (1.25,1.25) {$Q$};
      \node (Q1) at (3.75,1.25) {$\left|Q\right.$};
      \node (Q2) at (1.25,3.75) {$\underline{Q}$};
      \node (Q12) at (3.75,3.75) {$\left|\underline{Q}\right.$};
      \node (x) at (0,0) [circle,fill,label=45:{$x$}] {};
      \node (x1) at (2.5,0) [circle,fill,label=135:{$x_{1}$}] {};
      \node (x11) at (5,0) [circle,fill,label=135:{$x_{11}$}] {};
      \node (x2) at (0,2.5) [circle,fill,label=-45:{$x_{2}$}] {};
      \node (x12) at (2.5,2.5) [circle,fill,label=-135:{$x_{12}$}] {};
      \node (x112) at (5,2.5) [circle,fill,label=-135:{$x_{112}$}] {};
      \node (x22) at (0,5) [circle,fill,label=-45:{$x_{22}$}] {};
      \node (x122) at (2.5,5) [circle,fill,label=-135:{$x_{122}$}] {};
      \node (x1122) at (5,5) [circle,fill,label=-135:{$x_{1122}$}] {};
      \draw [ultra thick] (x) to node [swap] {$\alpha_{1}$} (x1) to node [swap] {$\alpha_{11}$} (x11);
      \draw [thin] (x2) to node {$\alpha_{1}$} (x12) to node {$\alpha_{11}$} (x112);
      \draw [ultra thick] (x1122) to node [swap] {$\alpha_{11}$} (x122) to node [swap] {$\alpha_{1}$} (x22);
      \draw [ultra thick] (x22) to node [swap] {$\alpha_{22}$} (x2) to node [swap] {$\alpha_{2}$} (x);
      \draw [thin] (x1) to node [swap] {$\alpha_{2}$} (x12) to node [swap] {$\alpha_{22}$} (x122);
      \draw [ultra thick] (x11) to node [swap] {$\alpha_{2}$} (x112) to node [swap] {$\alpha_{22}$} (x1122);
   \end{tikzpicture}
   }\qquad
   \subfloat[Translation of elementary cells]{\label{fig:6a}
   \begin{tikzpicture}[auto]
      \node at (0.75,0.75) {$Q$};
      \node at (3.75,0.75) {$Q$};
      \node at (2.25,0.75) {$\left|Q\right.$};
      \node at (5.25,0.75) {$\left|Q\right.$};
      \node at (0.75,2.25) {$\underline{Q}$};
      \node at (3.75,2.25) {$\underline{Q}$};
      \node at (2.25,2.25) {$\left|\underline{Q}\right.$};
      \node at (5.25,2.25) {$\left|\underline{Q}\right.$};
      \node at (0.75,3.75) {$Q$};
      \node at (3.75,3.75) {$Q$};
      \node at (2.25,3.75) {$\left|Q\right.$};
      \node at (5.25,3.75) {$\left|Q\right.$};
      \node at (0.75,5.25) {$\underline{Q}$};
      \node at (3.75,5.25) {$\underline{Q}$};
      \node at (2.25,5.25) {$\left|\underline{Q}\right.$};
      \node at (5.25,5.25) {$\left|\underline{Q}\right.$};
      \draw [ultra thick] (0,0) rectangle (6,6);
      \draw [thin] (0,1.5) -- (6,1.5);
      \draw [ultra thick] (0,3) -- (6,3);
      \draw [thin] (0,4.5) -- (6,4.5);
      \draw [thin] (1.5,0) -- (1.5,6);
      \draw [ultra thick] (3,0) -- (3,6);
      \draw [thin] (4.5,0) -- (4.5,6);
   \end{tikzpicture}
   }
   \caption{Embedding in $\Z^{2}$}
\end{figure}
Integrability of such a non-autonomous system of quad-equations understood as 3D consistency will be discussed at the end of the next section.
\begin{defi}[Asymmetric quad-equation]
An equation is called \emph{asymmetric} if $Q\neq\left|Q\right.$ or $Q\neq\underline{Q}$.
\end{defi}
A complete classification of multi-affine polynomials is given in \cite{classification}. We will now review the definitions, notations and results from that paper and \cite{lagrangian} which are necessary for the understanding of the present text.\par
For a multi-affine polynomial $Q\left(x_{1},x_{2},x_{3},x_{4}\right)$, one can apply certain discriminant-like operations to eliminate some of the variables $x_{1}$, $x_{2}$, $x_{3}$ and $x_{4}$.\par
Eliminating two out of four variables leads to a biquadratic polynomial. Thus, to any $Q$ there correspond six \emph{biquadratics}, assigned to the edges and diagonals of a quadrilateral. We distinguish between \emph{non-degenerate} and \emph{degenerate biquadratics}.\par
Eliminating three out of four variables in $Q$ leads to a quartic polynomial, called \emph{discriminant} and assigned to the remaining vertex of the quadrilateral.\par
Due to the classification, quad-equations can be divided in three groups:
\begin{itemize}
\item \emph{Type~Q} equations with all six biquadratics non-degenerate, 
\item \emph{type~\Hvier} equations with exactly two out of six biquadratics non-degenerate, and
\item \emph{type~\Hsechs} equations with all biquadratics degenerate.
\end{itemize}\par
\begin{figure}[htbp]
   \centering
   \subfloat[Quad-equation of type Q, all biquadratics non-de\-gen\-er\-ate]{\label{fig:3}
   \begin{tikzpicture}[auto]
      \node (x1) at (0,0) [circle,fill,label=-135:$x_{1}$] {};
      \node (x4) at (0,2.5) [circle,fill,label=135:$x_{4}$] {};
      \node (x2) at (2.5,0) [circle,fill,label=-45:$x_{2}$] {};
      \node (x3) at (2.5,2.5) [circle,fill,label=45:$x_{3}$] {};
      \draw [ultra thick] (x2) to node {$\alpha_{1}$} (x1);
      \draw [ultra thick] (x4) to node {$\alpha_{1}$} (x3);
      \draw [ultra thick] (x3) to node {$\alpha_{2}$} (x2);
      \draw [ultra thick] (x1) to node {$\alpha_{2}$} (x4);
      \draw [ultra thick] (x2) to (x4);
      \draw [ultra thick] (x1) to (x3);
   \end{tikzpicture}
   }\qquad
   \subfloat[Quad-equation of type~\Hsechs, all biquadratics degenerate]{\label{fig:4}
   \begin{tikzpicture}[auto]
      \node (x1) at (0,0) [circle,fill,gray,label=-135:$x_{1}$] {};
      \node (x4) at (0,2.5) [circle,fill,gray,label=135:$x_{4}$] {};
      \node (x2) at (2.5,0) [circle,fill,gray,label=-45:$x_{2}$] {};
      \node (x3) at (2.5,2.5) [circle,fill,gray,label=45:$x_{3}$] {};
      \draw [thin] (x2) to (x1);
      \draw [thin] (x4) to (x3);
      \draw [thin] (x3) to (x2);
      \draw [thin] (x1) to (x4);
      \draw [thin] (x2) to (x4);
      \draw [thin] (x1) to (x3);
   \end{tikzpicture}
   }
   \caption{Biquadratics patterns of type~Q and type~\Hsechs\ equations; non-degenerate biquadratics are indicated by thick lines}
\label{fig:bp}
\end{figure}
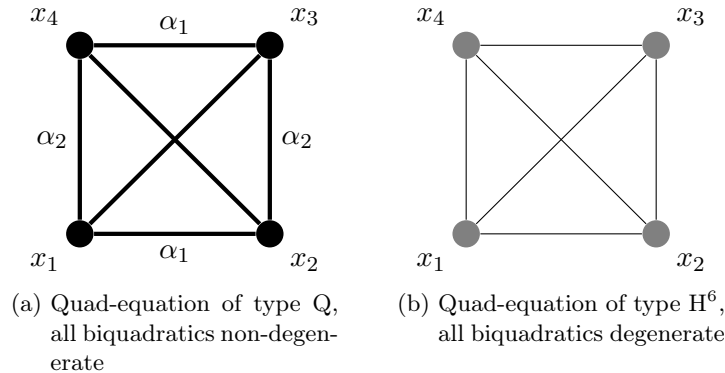

We introduced so-called \emph{biquadratics patterns}, which allow us to visually keep track of the distribution of the degenerate and non-degenerate biquadratics. Type~Q equations have one and the same biquadratics pattern as shown in Figure~\ref{fig:3}. Likewise, type~\Hsechs\ equations have also one and the same biquadratics pattern as shown in Figure~\ref{fig:4}.\par
For type~\Hvier\ equations, there are four degenerate and two non-degenerate biquadratics which allows two different patterns (up to rotations of a quadrilateral): a rhombic one, shown in Figure~\ref{fig:2a}, and a trapezoidal one, shown in Figure~\ref{fig:2b}.\par
Moreover, edges carrying non-degenerate biquadratics always connecting vertices with discriminants of the same equivalence class modulo M\"obius transformations. This need not to be true for edges carrying degenerate biquadratics. This fact leads to the black-and-white coloring of the vertices in the biquadratics patterns of type~\Hvier\ equations, where vertices with discriminants of the same equivalence class have the same color.\par
In the case of type~Q-equations all vertices carry discriminants of the same equivalence class (all six biquadratics are non-degenerate) and, therefore, they all have the same color which can be either black or white depending possibly on the coloring coming from the equations on a neighboring face if it happens to be of type~\Hvier. This is, e.g., the case for the cubes we will discuss in the next section.\par
Due to the absence of non-degenerate biquadratics for type~\Hsechs\ equations there does not exist a well-defined black-and-white coloring of vertices in this case considering only one quadrilateral. However, such a coloring can also be induced by the coloring of neighboring faces of a cube. We illustrate this fact by coloring these vertices gray.\par
The canonical forms $Q_{1}^{\epsilon}$ -- $Q_{3}^{\epsilon}$ (see \cite{lagrangian}) and $Q_{4}$ (see \cite{ABS2}) of type~Q equations and the canonical forms $H_{1}^{\epsilon}$ -- $H_{3}^{\epsilon}$ (see \cite{lagrangian}) depend on two parameters $\alpha_1,\alpha_2$ which can be assigned to the pairs of opposite edges of the elementary square.\par
Now, we should again have a look at the notion of an asymmetric quad-equation which is already used in the title. An asymmetric quad-equation is a quad-equation which is not equal to one the canonical forms $Q_{1}^{\epsilon}$ -- $Q_{3}^{\epsilon}$, $Q_{4}$ or $H_{1}^{0}$ -- $H_{3}^{0}$. Note, that in the last two instances, the parameter $\epsilon$ is set to $0$. In other words, asymmetric quad-equations can be, e.g., type~Q equations which are not in the canonical form or type~\Hvier\ equations in the canonical form with the parameter $\epsilon\neq0$.

\section{3D consistency, integrability, B\"acklund transformations} \label{cubes}

\begin{figure}[htbp]
   \centering
   \subfloat[A B\"acklund cube]{\label{fig:cube}
   \begin{tikzpicture}[auto,scale=0.8]
      \node (x) at (0,0) [circle,fill,label=-45:$x$] {};
      \node (x1) at (4,0) [circle,fill,label=-45:$x_{1}$] {};
      \node (x2) at (1.5,1.5) [circle,fill,label=-45:$x_{2}$] {};
      \node (x3) at (0,4) [circle,fill,label=-45:$x_{3}$] {};
      \node (x12) at (5.5,1.5) [circle,fill,label=-45:$x_{12}$] {};
      \node (x13) at (4,4) [circle,fill,label=-45:$x_{13}$] {};
      \node (x23) at (1.5,5.5) [circle,fill,label=-45:$x_{23}$] {};
      \node (x123) at (5.5,5.5) [circle,fill,label=-45:$x_{123}$] {};
      \node (A) at (2.75,0.75) {$A$};
      \node (Aq) at (2.75,4.75) {$D$};
      \node (B) at (0.75,2.75) {$B$};
      \node (Bq) at (4.75,2.75) {$E$};
      \node (C) at (2,2) {$C$};
      \node (Cq) at (3.5,3.5) {$F$};
      \draw (x) to (x1) to (x12) to (x123) to (x23) to (x3) to (x);
      \draw (x3) to (x13) to (x1);
      \draw (x13) to (x123);
      \draw [dashed] (x) to (x2) to (x12);
      \draw [dashed] (x2) to (x23);
      \draw [dotted,thick] (A) to (Aq);
      \draw [dotted,thick] (B) to (Bq);
      \draw [dotted,thick] (C) to (Cq);
   \end{tikzpicture}
   }\quad
   \subfloat[Making tetrahedra to faces (see Lem\-ma~\ref{tetrahedronUse})]{\label{fig:cube2}
   \begin{tikzpicture}[auto,scale=0.8]
      \node (x) at (0,0) [circle,fill,label=-45:$x$] {};
      \node (x1) at (4,0) [circle,fill,label=-45:$x_{13}$] {};
      \node (x2) at (1.5,1.5) [circle,fill,label=-45:$x_{23}$] {};
      \node (x3) at (0,4) [circle,fill,label=-45:$x_{3}$] {};
      \node (x12) at (5.5,1.5) [circle,fill,label=-45:$x_{12}$] {};
      \node (x13) at (4,4) [circle,fill,label=-45:$x_{1}$] {};
      \node (x23) at (1.5,5.5) [circle,fill,label=-45:$x_{2}$] {};
      \node (x123) at (5.5,5.5) [circle,fill,label=-45:$x_{123}$] {};
      \node (A) at (2.75,0.75) {$K$};
      \node (Aq) at (2.75,4.75) {$L$};
      \node (B) at (0.75,2.75) {$B$};
      \node (Bq) at (4.75,2.75) {$E$};
      \node (C) at (2,2) {$C$};
      \node (Cq) at (3.5,3.5) {$F$};
      \draw (x) to (x1) to (x12) to (x123) to (x23) to (x3) to (x);
      \draw (x3) to (x13) to (x1);
      \draw (x13) to (x123);
      \draw [dashed] (x) to (x2) to (x12);
      \draw [dashed] (x2) to (x23);
      \draw [dotted,thick] (A) to (Aq);
      \draw [dotted,thick] (B) to (Bq);
      \draw [dotted,thick] (C) to (Cq);
   \end{tikzpicture}
   }
   \caption{Equations on a Cube}
\end{figure}
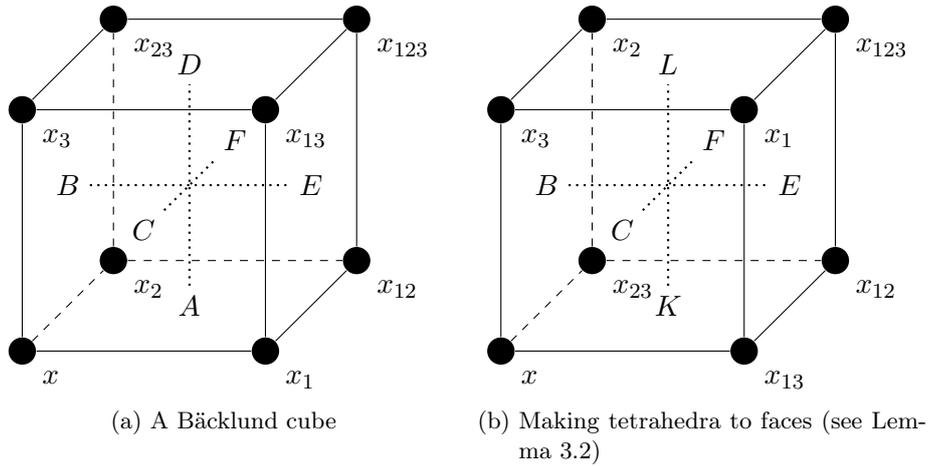

We will now consider six-tuples of (a priori different) quad-equations assigned to the faces of a 3D cube:
\begin{align}\label{system}
&A\left(x,x_{1},x_{2},x_{12}\right)=0,&
&D\left(x_{3},x_{13},x_{23},x_{123}\right)=0,\notag\\
&B\left(x,x_{2},x_{3},x_{23}\right)=0,&
&E\left(x_{1},x_{12},x_{13},x_{123}\right)=0,\\
&C\left(x,x_{1},x_{3},x_{13}\right)=0,&
&F\left(x_{2},x_{12},x_{23},x_{123}\right)=0,\notag
\end{align}
see Figure~\ref{fig:cube}. Such a six-tuple is \emph{3D consistent} (or a \emph{B\"acklund cube}) if, for arbitrary initial data $x$, $x_{1}$, $x_{2}$ and $x_{3}$,
the three values for $x_{123}$ (calculated by using $D=0$, $E=0$ or $F=0$) coincide. A 3D consistent six-tuple is said to possess the \emph{tetrahedron property} if there exist two polynomials $K$ and $L$ such that the equations
\begin{align*}
&K\left(x,x_{12},x_{13},x_{23}\right)=0,&
&L\left(x_{1},x_{2},x_{3},x_{123}\right)=0
\end{align*}
are satisfied for every solution of the six-tuple. It can be shown that the polynomials $K$ and $L$ are multi-affine and irreducible (see \cite{classification}).\par
\begin{figure}[htbp]
   \centering
   \subfloat[First case: all face e\-qua\-tions of type~\Hvier, tetrahedron equations of type Q]{\label{fig:bp1.1}
   \begin{tikzpicture}[auto]
      \node (x) at (0,0) [circle,fill] {};
      \node (x1) at (2,0) [circle,draw,ultra thick] {};
      \node (x3) at (0,2) [circle,draw,ultra thick] {};
      \node (x12) at (3,0.75) [circle,fill] {};
      \node (x13) at (2,2) [circle,fill] {};
      \node (x23) at (1,2.75) [circle,fill] {};
      \node (x123) at (3,2.75) [circle,draw,ultra thick] {};
      \draw [ultra thick] (x) to (x13) to (x12);
      \draw [ultra thick] (x13) to (x23);
      \draw [ultra thick] (x1) to (x3) to (x123) to (x1);
      \draw [thin] (x) to (x1) to (x12) to (x123) to (x23) to (x3) to (x);
      \draw [thin] (x1) to (x13) to (x3);
      \draw [thin] (x13) to (x123);
   \end{tikzpicture}
   }\qquad
   \subfloat[Second case: two pairs of face equations and tetrahedron equations of type~\Hvier, one pair of face equations of type~Q]{\label{fig:bp1.2}
   \begin{tikzpicture}[auto]
      \node (x) at (0,0) [circle,fill] {};
      \node (x1) at (2,0) [circle,fill] {};
      \node (x3) at (0,2) [circle,draw,ultra thick] {};
      \node (x12) at (3,0.75) [circle,fill] {};
      \node (x13) at (2,2) [circle,draw,ultra thick] {};
      \node (x23) at (1,2.75) [circle,draw,ultra thick] {};
      \node (x123) at (3,2.75) [circle,draw,ultra thick] {};
      \draw [ultra thick] (x) to (x1) to (x12);
      \draw [ultra thick] (x3) to (x13) to (x123) to (x23) to (x13);
      \draw [ultra thick] (x123) to (x3) to (x23);
      \draw [thin] (x) to (x3) to (x1) to (x123) to (x12);
      \draw [thin] (x) to (x13) to (x12);
      \draw [thin] (x1) to (x13);
   \end{tikzpicture}
   }\qquad
   \subfloat[Third case: all face and tetrahedron equations of type~\Hvier]{\label{fig:bp1.3}
   \begin{tikzpicture}[auto]
      \node (x) at (0,0) [circle,fill] {};
      \node (x1) at (2,0) [circle,draw,ultra thick] {};
      \node (x3) at (0,2) [circle,fill] {};
      \node (x12) at (3,0.75) [circle,fill] {};
      \node (x13) at (2,2) [circle,draw,ultra thick] {};
      \node (x23) at (1,2.75) [circle,draw,ultra thick] {};
      \node (x123) at (3,2.75) [circle,fill] {};
      \draw [ultra thick] (x) to (x3) to (x123) to (x12);
      \draw [ultra thick] (x1) to (x13) to (x23);
      \draw [thin] (x) to (x1) to (x12) to (x13) to (x);
      \draw [thin] (x3) to (x1) to (x123) to (x23) to (x3) to (x13) to (x123);
   \end{tikzpicture}
   }
   \caption{Biquadratics patterns containing type~\Hvier\ equations and not containing type~\Hsechs\ equations}
   \label{fig:bp1}
\end{figure}
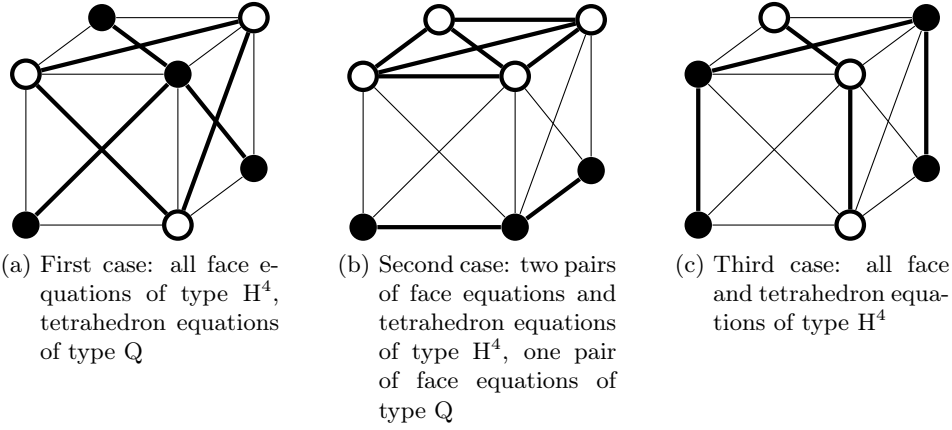

A complete classification of 3D consistent six-tuples~\eqref{system} possessing the tetrahedron property, modulo (possibly different) M\"obius transformations of the eight fields $x$, $x_{i}$, $x_{ij}$, $x_{123}$, is given in \cite{classification}. We will now mention the important general properties of such six-tuples:
\begin{lemma}\label{lem:opposite}
Equations on opposite faces are of the same type, and, in the case of type~\Hvier\ equations, corresponding edges both carry either non-degenerate or degenerate biquadratics. We say briefly that equations on opposite faces have the same biquadratics pattern.
\end{lemma}
We mostly consider B\"acklund cubes consisting of six equations of type~Q and those whose equations are not all of type~Q and which do not contain type~\Hsechs\ equations. Taking into account Lemma \ref{lem:opposite}, one easily sees that three different combinatorial arrangements of biquadratics patterns are possible, as indicated on Figure~\ref{fig:bp1}.
Arrangements on Figures~\ref{fig:bp1.1} and \ref{fig:bp1.2} are related to each other by the following construction, see \cite{classification} for a proof:
\begin{lemma} \label{tetrahedronUse}
Consider a 3D consistent six-tuple~\eqref{system} possessing the tetrahedron property expressed by the two equations
\begin{align*}
&K\left(x,x_{12},x_{13},x_{23}\right)=0,&
&L\left(x_{1},x_{2},x_{3},x_{123}\right)=0.
\end{align*}
Then the six-tuple
\begin{align}\label{dualsystem}
&K\left(x,x_{12},x_{13},x_{23}\right)=0,&
&L\left(x_{1},x_{2},x_{3},x_{123}\right)=0,\notag\\
&B\left(x,x_{2},x_{3},x_{23}\right)=0,&
&E\left(x_{1},x_{12},x_{13},x_{123}\right)=0,\\
&C\left(x,x_{1},x_{3},x_{13}\right)=0,&
&F\left(x_{2},x_{12},x_{23},x_{123}\right)=0,\notag
\end{align}
assigned to the faces of a 3D cube as on Figure~\ref{fig:cube2}, is 3D consistent and possesses the tetrahedron property expressed by the two equations
\begin{align*}
&A\left(x,x_{1},x_{2},x_{12}\right)=0,&
&D\left(x_{3},x_{13},x_{23},x_{123}\right)=0.
\end{align*}
3D consistency of \eqref{dualsystem} is understood as the property of the initial value problem with the initial data $x$, $x_{3}$, $x_{13}$ and $x_{23}$.
\end{lemma}
The way from six-tuple~\eqref{system} to six-tuple~\eqref{dualsystem} can be described by flipping the assignment to the vertices of $x_{1}$ and $x_{13}$ and furthermore of $x_{2}$ and $x_{23}$. This lemma suggests the consideration of the eight-tuple consisting of six face equations and two tetrahedron equations as a separate object.
\begin{defi}[Super-consistent eight-tuple on a decorated cube]
For a 3D consistent six-tuple of quad-equations possessing the tetrahedron property, the set consisting of the six face equations and the two tetrahedron equations is called a \emph{super-consistent eight-tuple} on a \emph{decorated cube}.
\end{defi}
The combinatorial structure behind a super-consistent eight-tuple, called a decorated cube in the above definition, consists of
\begin{itemize}
\item eight vertices of a cube carrying the fields,
\item twelve edges and twelve face diagonals carrying the parameters (if we consider tuples not containing equations of type \Hsechs) and
\item six faces and two tetrahedra carrying the equations.
\end{itemize}
Thus, it is richer than the standard 3D cube. The choice of a cube as a representative of a decorated cube can lead to different 3D systems (in a given system of coordinates), as exemplified in Lemma~\ref{tetrahedronUse}.\par
In Section~\ref{faces} we described how to construct a discrete integrable system from an arbitrary single quad-equation by the procedure of reflecting the squares (see Construction~\ref{const1}). We will now show how one can construct a B\"acklund transformation from one single B\"acklund cube by a reflecting procedure generalizing
(and subsuming) Construction~\ref{const1}. We will end up with a system of quad-equations on all faces of the lattice $\Z^{2}\times\left\{0,1\right\}$.
\begin{const}[Reflecting one B\"acklund cube]\label{const2}
We start with a B\"acklund cube as demonstrated in Figure~\ref{fig:3dcube3}. Then its neighbor in direction $1$ (in our picture the cube on the right hand side) will look like the cube in Figure~\ref{fig:3dcube4} in the notations of ``reflected quad-equations'' we introduced in Construction~\ref{const1}.\par
So, the procedure to get the neighboring cube, e.g., in direction $1$ is as follows: We take a copy of the first cube, reflect it in the plane of the 2D face $\left(x_{1},x_{12},x_{13},x_{123}\right)$ which is normal to direction $1$, and glue it to the original one along the common 2D face. Equations $E=0$ on the common face are the same, whereas the equations $\left|A\right.=0$, $\left|C\right.=0$, $\left|D\right.=0$ and $\left|F\right.=0$ on faces adjacent to the original cube are obtained by reflection of the equations $A=0$, $C=0$, $D=0$ and $F=0$, respectively. Finally, the polynomial $B$ is one and the same on both cubes.\par
In the same manner one can derive the neighbor of the original B\"acklund cube in direction $2$. Moreover, the common neighbor of the two considered neighbors of the original B\"acklund cube can also be obtained from them by the same reflection procedure. In fact, it can be easily seen that the construction is independent of the order of reflections (first in direction $1$ and then in direction $2$ or vice versa): For the equations on faces in the ``horizontal'' planes $\Z^{2}\times\left\{0\right\}$ and $\Z^{2}\times\left\{1\right\}$ this follows immediately from the considerations in Construction~\ref{const1}. On equations on ``vertical'' faces, e.g., the equation $C\left(x,x_{1},x_{3},x_{13}\right)=0$ the reflection procedures act as follows: Reflection in direction $1$ leads to the equation $\left|C\right.\left(x_{1},x_{11},x_{13},x_{113}\right)=0$, and then reflection in direction $2$ leads to the equation $\left|C\right.\left(x_{122},x_{1122},x_{1223},x_{11223}\right)=0$. On the other hand, the reflection of $C\left(x,x_{1},x_{3},x_{13}\right)=0$ in direction $2$ leads to the equations $C\left(x_{22},x_{122},x_{223},x_{1223}\right)=0$, and then the reflection in direction $1$ leads to the equation $\left|C\right.\left(x_{122},x_{1122},x_{1223},x_{11223}\right)=0$ which is the same as for the first order of reflections.\par
The whole lattice $\Z^{2}\times\left\{0,1\right\}$ can be filled by translating the elementary $2\times 2$-cell containing all four B\"acklund cubes we considered above.
\end{const}
\begin{rem}
Observe that the equations in the plane $\Z\times\left\{0\right\}$ are obtained from the equation $A=0$ according to Construction~\ref{const1}. Likewise, the equations in the plane $\Z\times\left\{1\right\}$ are obtained from the equation $D=0$ according to Construction~\ref{const1}.
\end{rem}
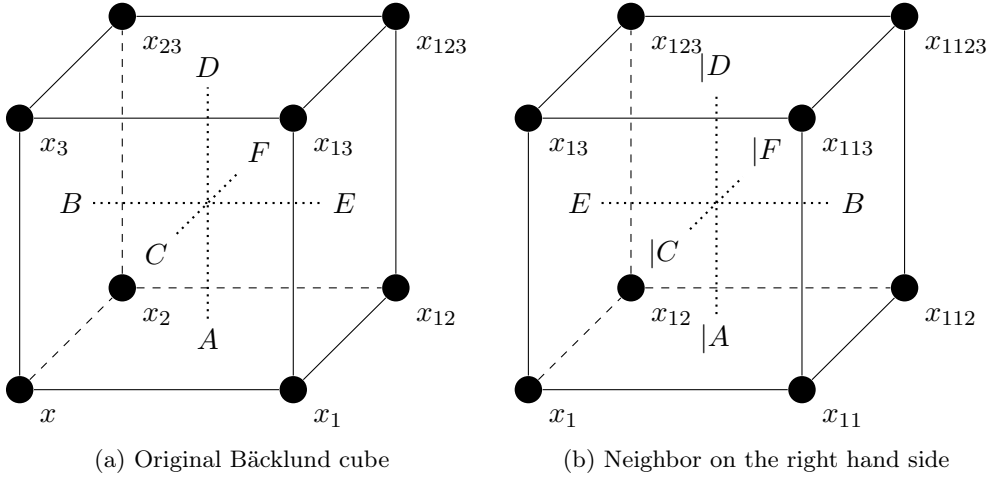
\begin{figure}[htbp]
   \centering
   \subfloat[Original B\"acklund cube]{\label{fig:3dcube3}
   \begin{tikzpicture}[auto,scale=0.9]
      \node (x) at (0,0) [circle,fill,label=-45:$x$] {};
      \node (x1) at (4,0) [circle,fill,label=-45:$x_{1}$] {};
      \node (x2) at (1.5,1.5) [circle,fill,label=-45:$x_{2}$] {};
      \node (x3) at (0,4) [circle,fill,label=-45:$x_{3}$] {};
      \node (x12) at (5.5,1.5) [circle,fill,label=-45:$x_{12}$] {};
      \node (x13) at (4,4) [circle,fill,label=-45:$x_{13}$] {};
      \node (x23) at (1.5,5.5) [circle,fill,label=-45:$x_{23}$] {};
      \node (x123) at (5.5,5.5) [circle,fill,label=-45:$x_{123}$] {};
      \node (A) at (2.75,0.75) {$A$};
      \node (Aq) at (2.75,4.75) {$D$};
      \node (B) at (0.75,2.75) {$B$};
      \node (Bq) at (4.75,2.75) {$E$};
      \node (C) at (2,2) {$C$};
      \node (Cq) at (3.5,3.5) {$F$};
      \draw (x) to (x1) to (x12) to (x123) to (x23) to (x3) to (x);
      \draw (x3) to (x13) to (x1);
      \draw (x13) to (x123);
      \draw [dashed] (x) to (x2) to (x12);
      \draw [dashed] (x2) to (x23);
      \draw [dotted,thick] (A) to (Aq);
      \draw [dotted,thick] (B) to (Bq);
      \draw [dotted,thick] (C) to (Cq);
   \end{tikzpicture}
   }\hspace{1mm}
   \subfloat[Neighbor on the right hand side]{\label{fig:3dcube4}
   \begin{tikzpicture}[auto,scale=0.9]
      \node (x) at (0,0) [circle,fill,label=-45:$x_{1}$] {};
      \node (x1) at (4,0) [circle,fill,label=-45:$x_{11}$] {};
      \node (x2) at (1.5,1.5) [circle,fill,label=-45:$x_{12}$] {};
      \node (x3) at (0,4) [circle,fill,label=-45:$x_{13}$] {};
      \node (x12) at (5.5,1.5) [circle,fill,label=-45:$x_{112}$] {};
      \node (x13) at (4,4) [circle,fill,label=-45:$x_{113}$] {};
      \node (x23) at (1.5,5.5) [circle,fill,label=-45:$x_{123}$] {};
      \node (x123) at (5.5,5.5) [circle,fill,label=-45:$x_{1123}$] {};
      \node (A) at (2.75,0.75) {$\left|A\right.$};
      \node (Aq) at (2.75,4.75) {$\left|D\right.$};
      \node (B) at (0.75,2.75) {$E$};
      \node (Bq) at (4.75,2.75) {$B$};
      \node (C) at (2,2) {$\left|C\right.$};
      \node (Cq) at (3.5,3.5) {$\left|F\right.$};
      \draw (x) to (x1) to (x12) to (x123) to (x23) to (x3) to (x);
      \draw (x3) to (x13) to (x1);
      \draw (x13) to (x123);
      \draw [dashed] (x) to (x2) to (x12);
      \draw [dashed] (x2) to (x23);
      \draw [dotted,thick] (A) to (Aq);
      \draw [dotted,thick] (B) to (Bq);
      \draw [dotted,thick] (C) to (Cq);
   \end{tikzpicture}
   }
   \caption{Construction of a B\"acklund transformation from one single B\"acklund cube}
\end{figure}
This procedure allows us to define B\"acklund transformations:
\begin{defi}[B\"acklund transformation $x\overset{\lambda}{\mapsto}y$]\label{def:Baecklund}
Consider a system of quad-equations on the lattice $\Z^{2}\times\left\{0,1\right\}$ generated by the reflection procedure described in Construction~\ref{const2} and a solution $x$ of the system composed of the equations $A=0$, $\left|A\right.=0$, $\underline{A}=0$ and $\left|\underline{A}\right.=0$ on the ``horizontal'' sublattice $\Z^{2}\times\left\{0\right\}$.\par
Then we say that a solution $y$ of the system composed of the equations $D=0$, $\left|D\right.=0$, $\underline{D}=0$ and $\left|\underline{D}\right.=0$ on the sublattice $\Z^{2}\times\left\{1\right\}$ is a \emph{B\"acklund transform} of $x$ if the combination of $x$ and $y$ build a solution of the system on the whole lattice $\Z^{2}\times\left\{0,1\right\}$, i.e., also the equations on ``vertical'' faces are satisfied.\par
In this case, the B\"acklund parameter $\lambda$ is the parameter which is assigned to the vertical edges. The B\"acklund transformation $x\overset{\lambda}{\mapsto} y$ itself can be described by equations $B=0$, $E=0$, $C=0$, $F=0$ on the vertical faces of an elementary cube. Therefore, a B\"acklund transformation can be seen as one layer of B\"acklund cubes in the three-dimensional lattice $\Z^{3}$ (see Figure~\ref{fig:cube3}).
\end{defi}
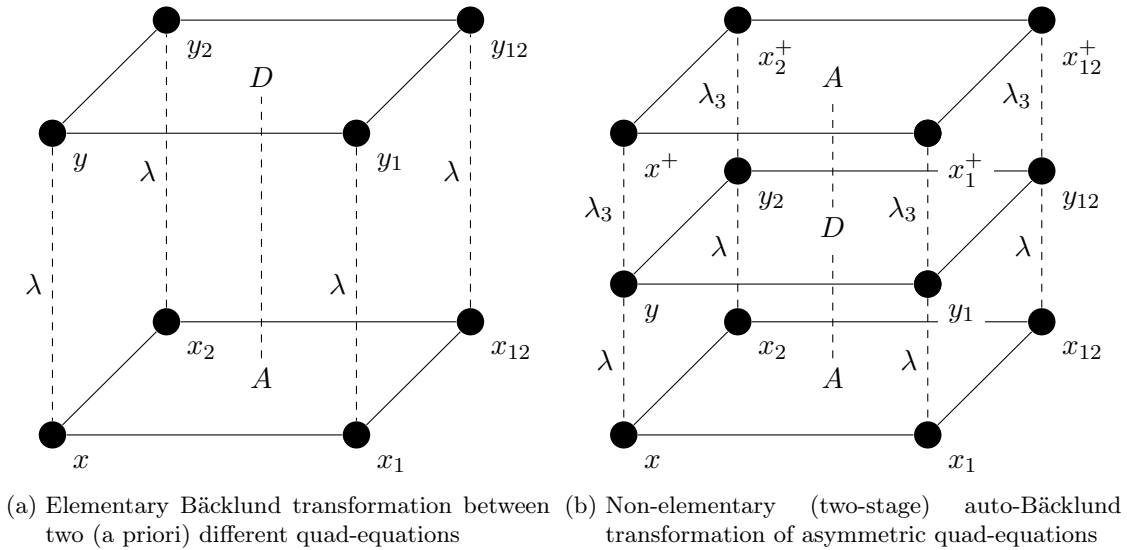
\begin{figure}[htbp]
   \centering
   \subfloat[Elementary B\"acklund transformation between two (a priori) different quad-equations]{\label{fig:cube3}
   \begin{tikzpicture}[auto]
      \node (x) at (0,0) [circle,fill,label=-45:$x$] {};
      \node (x1) at (4,0) [circle,fill,label=-45:$x_{1}$] {};
      \node (x2) at (1.5,1.5) [circle,fill,label=-45:$x_{2}$] {};
      \node (x3) at (0,4) [circle,fill,label=-45:$y$] {};
      \node (x12) at (5.5,1.5) [circle,fill,label=-45:$x_{12}$] {};
      \node (x13) at (4,4) [circle,fill,label=-45:$y_{1}$] {};
      \node (x23) at (1.5,5.5) [circle,fill,label=-45:$y_{2}$] {};
      \node (x123) at (5.5,5.5) [circle,fill,label=-45:$y_{12}$] {};
      \node (A) at (2.75,0.75) {$A$};
      \node (Aq) at (2.75,4.75) {$D$};
      \draw (x) to (x1);
      \draw (x1) to (x12);
      \draw [dashed] (x12) to node {$\lambda$} (x123);
      \draw (x123) to (x23);
      \draw (x23) to (x3);
      \draw [dashed] (x3) to node [swap] {$\lambda$} (x);
      \draw (x3) to (x13);
      \draw [dashed] (x13) to node [swap] {$\lambda$} (x1);
      \draw (x13) to (x123);
      \draw (x) to (x2);
      \draw (x2) to (x12);
      \draw [dashed] (x2) to node {$\lambda$} (x23);
      \draw [dashed] (A) to (Aq);
   \end{tikzpicture}
   }\hspace{0.5mm}
   \subfloat[Non-elementary (two-stage) auto-B\"acklund trans\-formation of asymmetric quad-equations]{\label{fig:cube4}
   \begin{tikzpicture}[auto]
      \node (x) at (0,0) [circle,fill,label=-45:$x$] {};
      \node (x1) at (4,0) [circle,fill,label=-45:$x_{1}$] {};
      \node (x2) at (1.5,1.5) [circle,fill,label=-45:$x_{2}$] {};
      \node (x3) at (0,4) [circle,fill,label=-45:$x^{+}$] {};
      \node (x12) at (5.5,1.5) [circle,fill,label=-45:$x_{12}$] {};
      \node (x13) at (4,4) [circle,fill,label=-45:$x_{1}^{+}$] {};
      \node (x23) at (1.5,5.5) [circle,fill,label=-45:$x_{2}^{+}$] {};
      \node (x123) at (5.5,5.5) [circle,fill,label=-45:$x_{12}^{+}$] {};
      \node (y) at (0,2) [circle,fill,label=-45:$y$] {};
      \node (y1) at (4,2) [circle,fill,label={[fill=white]-45:$y_{1}$}] {};
      \node (y2) at (1.5,3.5) [circle,fill,label=-45:$y_{2}$] {};
      \node (y12) at (5.5,3.5) [circle,fill,label=-45:$y_{12}$] {};
      \node (A) at (2.75,0.75) {$A$};
      \node (Aq) at (2.75,4.75) {$A$};
      \node (Q) at (2.75,2.75) {$D$};
      \draw (x) to (x1) to (x12);
      \draw [dashed] (x12) to node {$\lambda$} (y12) to node {$\lambda_{3}$} (x123);
      \draw (x123) to (x23) to (x3);
      \draw [dashed] (x3) to node [swap] {$\lambda_{3}$} (y) to node [swap] {$\lambda$} (x);
      \draw (x3) to (x13);
      \draw [dashed] (x13) to node [swap] {$\lambda_{3}$} (y1) to node [swap] {$\lambda$} (x1);
      \draw (x13) to (x123);
      \draw (y) to (y1) to (y12);
      \draw (x) to (x2) to (x12);
      \draw [dashed] (x2) to node {$\lambda$} (y2) to node {$\lambda_{3}$} (x23);
      \draw (y) to (y2) to (y12);
      \draw [dashed] (A) to (Q) to (Aq);
      \node (y1) at (4,2) [circle,fill,label={[fill=white]-45:$y_{1}$}] {};
      \node (x13) at (4,4) [circle,fill,label={[fill=white]-45:$x_{1}^{+}$}] {};
   \end{tikzpicture}
   }
   \caption{B\"acklund transformations}
\end{figure}
The main conceptual message of \cite{quadgraphs,ABS1} is that 3D consistency is synonymous with integrability. In the situations considered there, where equations on opposite faces of the cube are shifted versions of one another, it was demonstrated how to derive (auto-)B\"acklund transformations and zero curvature representations from a 3D consistent system.\par
It might be not immediately obvious whether these integrability attributes of 2D systems with an elementary $2\times 2$ building-block can still be derived from our 3D systems, where the equations on opposite faces of one elementary cube happen to be different. But this is the case, indeed: One can find zero curvature representations (see \cite{classification} for more details) and (properly generalized) auto-B\"acklund transformations for these systems.\par
In the symmetric case, i.e., for the systems from the ABS-list, the above defined elementary B\"acklund transformations (see Figure~\ref{fig:cube3}) are auto-B\"acklund transformations by themselves because in this case, the polynomials $A$ and $D$ coincide. A more detailed demonstration of this situation can be found for example in \cite{ddg}.\par
In the case of asymmetric quad-equations, we have to consider a picture which is a little bit more extensive (see Figure~\ref{fig:cube4}) in order to get (non-elementary) auto-B\"acklund transformations. In this case an auto-B\"acklund transformation can be seen as two layers of B\"acklund cubes. 
\begin{const}[Two layers of B\"acklund cubes]\label{const3}
Consider a layer of B\"acklund cubes, i.e., a 3D consistent system of quad-equations on $\Z^{2}\times\left\{0,1\right\}$ generated by the reflection procedure described in Construction~\ref{const2}.\par
The second layer can be derived from the original one on the lattice $\Z^{2}\times\left\{0,1\right\}$ by reflecting the B\"acklund cubes in the plane $\Z^{2}\times\left\{1\right\}$ in order to get a 3D consistent system of quad-equations on the lattice $\Z^{2}\times\left\{0,1,2\right\}$.
\end{const}
Moreover, repeatedly reflecting cubes in horizontal planes leads to a 3D consistent system of quad-equations on the whole lattice $\Z^{3}$.\par
\begin{const}[Two-stage auto-B\"acklund transformation $x\overset{\lambda}{\mapsto} y\overset{\lambda_{3}}{\mapsto} x^{+}$]\label{th:autoBaeck}
Consider a 3D consistent system of quad-equations on the lattice $\Z^{2}\times\left\{0,1,2\right\}$ generated by the reflection procedure described in Construction~\ref{const3}. Furthermore, consider a solution $x$ of the system composed of the equations $A=0$, $\left|A\right.=0$, $\underline{A}=0$ and $\left|\underline{A}\right.=0$ on the sublattice $\Z^{2}\times\left\{0\right\}$ and one of its B\"acklund transforms $y$ satisfying the system composed of the equations $D=0$, $\left|D\right.=0$, $\underline{D}=0$ and $\left|\underline{D}\right.=0$ on the sublattice $\Z^{2}\times\left\{1\right\}$ with B\"acklund parameter $\lambda$.\par
Then a B\"acklund transform $x^{+}$ of $y$ satisfying the system composed of the equations $A=0$, $\left|A\right.=0$, $\underline{A}=0$ and $\left|\underline{A}\right.=0$ on the sublattice $\Z^{2}\times\left\{2\right\}$ with B\"acklund parameter $\lambda_{3}$ is a (non-elementary) auto-B\"acklund transform of $x$.
\end{const}

\section{4D consistent systems} \label{consi}
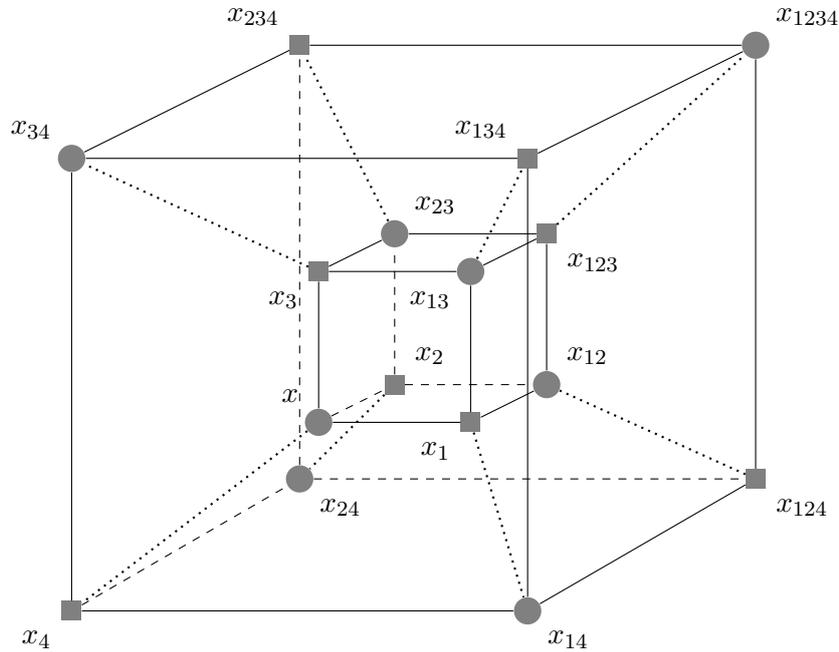
\begin{figure}[htbp]
   \centering
   \begin{tikzpicture}
      \node (x) at (0,0) [circle,fill=gray,label=135:$x$] {};
      \node (x1) at (2,0) [fill=gray,label=-135:$x_{1}$] {};
      \node (x2) at (1,0.5) [fill=gray,label=45:$x_{2}$] {};
      \node (x3) at (0,2) [fill=gray,label=-135:$x_{3}$] {};
      \node (x4) at (-3.25,-2.5) [fill=gray,label=-135:$x_{4}$] {};
      \node (x12) at (3,0.5) [circle,fill=gray,label=45:$x_{12}$] {};
      \node (x13) at (2,2) [circle,fill=gray,label=-135:$x_{13}$] {};
      \node (x14) at (2.75,-2.5) [circle,fill=gray,label=-45:$x_{14}$] {};
      \node (x23) at (1,2.5) [circle,fill=gray,label=45:$x_{23}$] {};
      \node (x24) at (-0.25,-0.75) [circle,fill=gray,label=-45:$x_{24}$] {};
      \node (x34) at (-3.25,3.5) [circle,fill=gray,label=135:$x_{34}$] {};
      \node (x123) at (3,2.5) [fill=gray,label=-45:$x_{123}$] {};
      \node (x124) at (5.75,-0.75) [fill=gray,label=-45:$x_{124}$] {};
      \node (x134) at (2.75,3.5) [fill=gray,label=135:$x_{134}$] {};
      \node (x234) at (-0.25,5) [fill=gray,label=135:$x_{234}$] {};
      \node (x1234) at (5.75,5) [circle,fill=gray,label=45:$x_{1234}$] {};
      \draw (x) to (x1) to (x12) to (x123) to (x23) to (x3) to (x);
      \draw (x4) to (x14) to (x124) to (x1234) to (x234) to (x34) to (x4);
      \draw (x1) to (x13) to (x3);
      \draw (x14) to (x134) to (x34);
      \draw (x13) to (x123);
      \draw (x134) to (x1234);
      \draw [dashed] (x) to (x2) to (x12);
      \draw [dashed] (x4) to (x24) to (x124);
      \draw [dashed] (x2) to (x23);
      \draw [dashed] (x24) to (x234);
      \draw [dotted,thick] (x) to (x4);
      \draw [dotted,thick] (x1) to (x14);
      \draw [dotted,thick] (x2) to (x24);
      \draw [dotted,thick] (x3) to (x34);
      \draw [dotted,thick] (x12) to (x124);
      \draw [dotted,thick] (x13) to (x134);
      \draw [dotted,thick] (x23) to (x234);
      \draw [dotted,thick] (x123) to (x1234);
   \end{tikzpicture}
   \caption{A Bianchi cube}
   \label{fig:4D cube}
\end{figure}
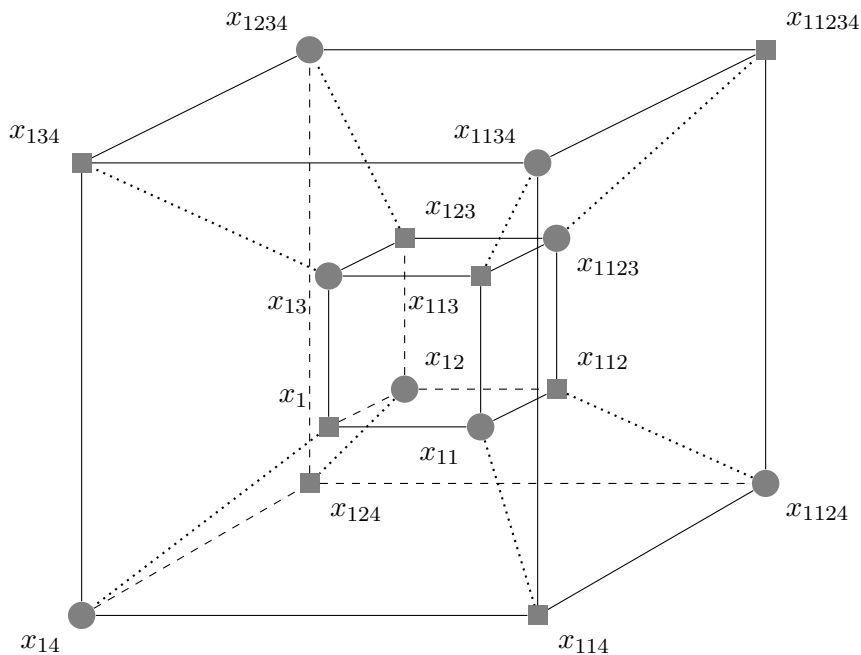
\begin{figure}[htbp]
   \centering
   \begin{tikzpicture}
      \node (x) at (0,0) [fill=gray,label=135:$x_{1}$] {};
      \node (x1) at (2,0) [circle,fill=gray,label=-135:$x_{11}$] {};
      \node (x2) at (1,0.5) [circle,fill=gray,label=45:$x_{12}$] {};
      \node (x3) at (0,2) [circle,fill=gray,label=-135:$x_{13}$] {};
      \node (x4) at (-3.25,-2.5) [circle,fill=gray,label=-135:$x_{14}$] {};
      \node (x12) at (3,0.5) [fill=gray,label=45:$x_{112}$] {};
      \node (x13) at (2,2) [fill=gray,label=-135:$x_{113}$] {};
      \node (x14) at (2.75,-2.5) [fill=gray,label=-45:$x_{114}$] {};
      \node (x23) at (1,2.5) [fill=gray,label=45:$x_{123}$] {};
      \node (x24) at (-0.25,-0.75) [fill=gray,label=-45:$x_{124}$] {};
      \node (x34) at (-3.25,3.5) [fill=gray,label=135:$x_{134}$] {};
      \node (x123) at (3,2.5) [circle,fill=gray,label=-45:$x_{1123}$] {};
      \node (x124) at (5.75,-0.75) [circle,fill=gray,label=-45:$x_{1124}$] {};
      \node (x134) at (2.75,3.5) [circle,fill=gray,label=135:$x_{1134}$] {};
      \node (x234) at (-0.25,5) [circle,fill=gray,label=135:$x_{1234}$] {};
      \node (x1234) at (5.75,5) [fill=gray,label=45:$x_{11234}$] {};
      \draw (x) to (x1) to (x12) to (x123) to (x23) to (x3) to (x);
      \draw (x4) to (x14) to (x124) to (x1234) to (x234) to (x34) to (x4);
      \draw (x1) to (x13) to (x3);
      \draw (x14) to (x134) to (x34);
      \draw (x13) to (x123);
      \draw (x134) to (x1234);
      \draw [dashed] (x) to (x2) to (x12);
      \draw [dashed] (x4) to (x24) to (x124);
      \draw [dashed] (x2) to (x23);
      \draw [dashed] (x24) to (x234);
      \draw [dotted,thick] (x) to (x4);
      \draw [dotted,thick] (x1) to (x14);
      \draw [dotted,thick] (x2) to (x24);
      \draw [dotted,thick] (x3) to (x34);
      \draw [dotted,thick] (x12) to (x124);
      \draw [dotted,thick] (x13) to (x134);
      \draw [dotted,thick] (x23) to (x234);
      \draw [dotted,thick] (x123) to (x1234);
   \end{tikzpicture}
   \caption{Bianchi cube on the 4D cube neighboring the Bianchi cube in Figure~\ref{fig:4D cube} in direction $1$}
   \label{fig:4D cube1}
\end{figure}

A 24-tuple of quad-equations on a 4D cube as shown in Figure~\ref{fig:4D cube} is \emph{4D consistent} or a \emph{Bianchi cube} if the six-tuples on all 3D facets are 3D consistent and the four values for $x_{1234}$ (calculated by using the six-tuples of 3D facets adjacent to $x_{1234}$) coincide for arbitrary initial data $x$, $x_{1}$, $x_{2}$, $x_{3}$ and $x_{4}$.
\begin{theo} \label{cons} Consider a 24-tuple of quad-equations on a 4D cube and suppose that all eight six-tuples of quad-equations corresponding to 3D facets are 3D consistent. Then it is 4D consistent.\par
\begin{proof}
Consider the initial value problem with initial data $x$, $x_{1}$, $x_{2}$, $x_{3}$ and $x_{4}$. Then we get unique values for $x_{12}$, $x_{13}$, $x_{14}$, $x_{23}$, $x_{24}$ and $x_{34}$ using the quad-equations on the corresponding faces adjacent to $x$. Then using the 3D consistency of the six-tuples on the 3D facets adjacent to $x$ one gets unique value for $x_{123}$, $x_{124}$, $x_{134}$ and $x_{234}$. Furthermore, using the 3D consistency of six-tuples on the 3D facets adjacent to $x_{1234}$ one gets four a priori different values for $x_{1234}$. However, since every two of these 3D facets have one face in common which is adjacent to $x_{1234}$, and therefore all these six-tuples have one equation in common which is dependent on $x_{1234}$, all these four values for $x_{1234}$ coincide. Therefore, the 24-tuple is 4D consistent.
\end{proof}
\end{theo}
\begin{rem}
This is an adaptation of a proof from \cite{ddg} that 4D consistent 3D systems are 5D consistent.
\end{rem}
The next step is to extend a Bianchi cube to a system of quad-equations on a 4D lattice:
\begin{const}[Reflecting one Bianchi cube]\label{const4}
We start with a Bianchi cube as dem\-onstrated in Figure~\ref{fig:4D cube}. Then the procedure to get the neighboring cube, e.g., in direction $1$ (see Figure~\ref{fig:4D cube1}) is as follows: We take a copy of the first 4D cube, reflect it in the 3D subspace of the 3D facet $\left(x_{1},x_{12},x_{13},x_{14},x_{123},x_{124},x_{134},x_{1234}\right)$ and glue it to the original one along the common 3D facet. So, the six equations on the common 3D facet (i.e. the equations on the right 3D facet in Figure~\ref{fig:4D cube} and on the left facet in Figure~\ref{fig:4D cube1}) are the same, the twelve equations on the 2D faces adjacent to the original 4D cube are obtained by reflection of the corresponding equations in the original Bianchi cube, and the polynomials of the six equations on the 2D faces not adjacent to the original 4D cube are the same as the ones of the equations on corresponding 2D faces.\par
In the same manner one can derive the neighbor of the original Bianchi cube in direction~$2$. Moreover, the common neighbor of the two considered neighbors of the original Bianchi cube can also be obtained from them by the same reflection procedure. Analogously to the considerations in Construction~\ref{const2}, one can show that the construction is independent of the order of reflections. The whole lattice $\Z^{2}\times\left\{0,1\right\}\times\left\{0,1\right\}$ can be filled by translating the elementary $2\times2$-cell containing all four Bianchi cubes we considered above. We call the resulting system a \emph{layer of Bianchi cubes}.\par
\end{const}
\begin{rem}
Observe that the equations in the subspaces $\Z^{2}\times\left\{0\right\}\times\left\{0,1\right\}$, $\Z^{2}\times\left\{1\right\}\times\left\{0,1\right\}$, $\Z^{2}\times\left\{0,1\right\}\times\left\{0\right\}$ and $\Z^{2}\times\left\{0,1\right\}\times\left\{1\right\}$ are obtained from corresponding B\"acklund cubes according to Construction~\ref{const2}.
\end{rem}
Moreover, repeatedly reflecting cubes in subspaces of types $\Z^{2}\times\left\{i\right\}\times\left\{j,j+1\right\}$, $\Z^{2}\times\left\{i,i+1\right\}\times\left\{j\right\}$ leads to a 4D consistent system of quad-equations on the whole lattice~$\Z^{4}$.

\section{Bianchi permutability of \texorpdfstring{B\"acklund}{Baecklund} transformations} \label{classi}

From now on, we only consider the situation that all quad-equations are of type~Q or type~\Hvier, because type~\Hsechs\ equations have less structure and are, moreover, less important. Then all 3D cubes we consider have biquadratics patterns as mentioned in Figure~\ref{fig:bp1} or carry only equations of type~Q. Furthermore, we restrict ourselves to the cases, were all B\"acklund cubes possesses the tetrahedron property. This allow us to make use of the results of \cite{classification} which are recapitulated at the beginning of Section~\ref{cubes}. Those lead us to the following lemma:
\begin{lemma}
Consider a Bianchi cube on a 4D cube and suppose that all eight B\"acklund cubes on 3D facets possess the tetrahedron property. Then the 16 tetrahedron equations build two super-consistent eight-tuples $\T$ and $\bar\T$ as demonstrated in Figure~\ref{fig:tetras}, where $\T$ depends on the fields on vertices whose indices have an even number of digits and $\bar{\T}$ depends on fields on vertices whose indices have an an odd number of digits.
\begin{proof}
We will first prove the super-consistency of $\T$. Consider the initial value problem with initial data $x$, $x_{12}$, $x_{13}$ and $x_{14}$. Then we get unique values for $x_{23}$, $x_{24}$ and $x_{34}$ using the tetrahedron equations $K\left(x,x_{1i},x_{1j},x_{ij}\right)=0$ with $i,j\in\left\{2,3,4\right\}$. Then using the tetrahedron equations $K\left(x_{1i},x_{ij},x_{ik},x_{1234}\right)=0$ with $i\in\left\{2,3,4\right\}$ and $j,k\in\left\{2,3,4\right\}\setminus\left\{i\right\}$ one gets three a priori different values for $x_{1234}$. Because of the 4D consistency of the original 24-tuple of quad-equations these three values must coincide. Therefore, the least mentioned six equations build a 3D consistent six-tuple with tetrahedron equations $K\left(x,x_{23},x_{24},x_{34}\right)=0$ and $K\left(x_{12},x_{13},x_{14},x_{1234}\right)=0$.\par Analogously, one can proof the super-consistency of $\bar{\T}$.
\end{proof}
\end{lemma}
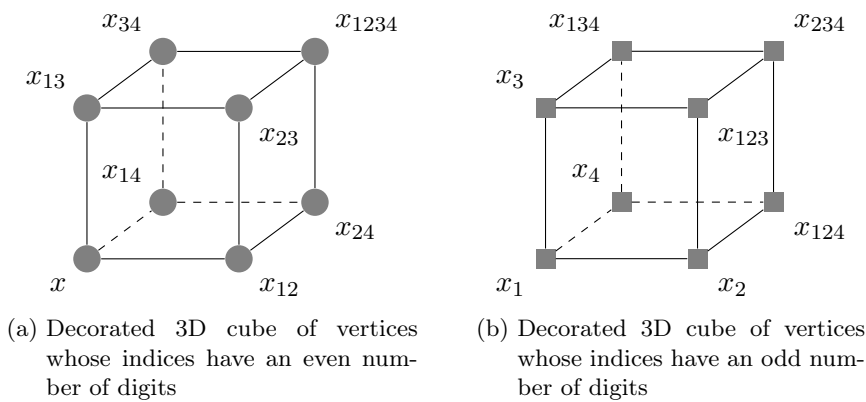
\begin{figure}[htbp]
   \centering
   \subfloat[Decorated 3D cube of vertices whose indices have an even number of digits]{\label{fig:a}
   \begin{tikzpicture}
      \node (x) at (0,0) [circle,fill,gray,label=-135:$x$] {};
      \node (x1) at (2,0) [circle,fill,gray,label=-45:$x_{12}$] {};
      \node (x2) at (1,0.75) [circle,fill,gray,label=135:$x_{14}$] {};
      \node (x12) at (3,0.75) [circle,fill,gray,label=-45:$x_{24}$] {};
      \node (x3) at (0,2) [circle,fill,gray,label=135:$x_{13}$] {};
      \node (x13) at (2,2) [circle,fill,gray,label=-45:$x_{23}$] {};
      \node (x23) at (1,2.75) [circle,fill,gray,label=135:$x_{34}$] {};
      \node (x123) at (3,2.75) [circle,fill,gray,label=45:$x_{1234}$] {};
      \draw (x) to (x1) to (x12) to (x123) to (x23) to (x3) to (x);
      \draw (x1) to (x13) to (x3);
      \draw (x13) to (x123);
      \draw [dashed] (x) to (x2) to (x12);
      \draw [dashed] (x2) to (x23);
   \end{tikzpicture}
   }\qquad
   \subfloat[Decorated 3D cube of vertices whose indices have an odd number of digits]{\label{fig:b}
   \begin{tikzpicture}
      \node (x) at (0,0) [fill,gray,label=-135:$x_{1}$] {};
      \node (x1) at (2,0) [fill,gray,label=-45:$x_{2}$] {};
      \node (x2) at (1,0.75) [fill,gray,label=135:$x_{4}$] {};
      \node (x12) at (3,0.75) [fill,gray,label=-45:$x_{124}$] {};
      \node (x3) at (0,2) [fill,gray,label=135:$x_{3}$] {};
      \node (x13) at (2,2) [fill,gray,label=-45:$x_{123}$] {};
      \node (x23) at (1,2.75) [fill,gray,label=135:$x_{134}$] {};
      \node (x123) at (3,2.75) [fill,gray,label=45:$x_{234}$] {};
      \draw (x) to (x1) to (x12) to (x123) to (x23) to (x3) to (x);
      \draw (x1) to (x13) to (x3);
      \draw (x13) to (x123);
      \draw [dashed] (x) to (x2) to (x12);
      \draw [dashed] (x2) to (x23);
   \end{tikzpicture}
   }
   \caption{Super-consistent eight-tuples composed from tetrahedron equations of the 3D facets of a 4D consistent 24-tuple}
   \label{fig:tetras}
\end{figure}
Now, we are able to proof Theorem~\ref{th:Bianchi} which we will repeat here for a better readability:
\begin{theo}\label{th:bianchi}
Given two B\"acklund cubes sharing one quad-equation, there exists an extension to a Bianchi cube. This extension is unique up to M\"obius transformations of fields not belonging to the original B\"acklund cubes (the fields $X$, $X_{1}$, $X_{2}$ and $X_{12}$ in Figure~\ref{fig:H1Super}).
\begin{proof}
Given two B\"acklund cubes sharing one equation $A\left(x,x_{1},x_{2},x_{12}\right)=0$. Then for every of these B\"acklund cubes one of its tetrahedron equations can be find on a face of the cube in Figure~\ref{fig:a} neighboring the edge $\left(x,x_{12}\right)$ and one can be find on a face of the cube in Figure~\ref{fig:b} neighboring the edge $\left(x_{1},x_{2}\right)$. Therefore, we know the biquadratics pattern and the black-and-white coloring of vertices for two neighboring faces of every of the cubed in Figures~\ref{fig:a} and \ref{fig:b} which defines the whole biquadratics pattern and coloring of vertices for both cubes. From this and the biquadratics patterns of the original B\"acklund cubes we get the whole biquadratics pattern and coloring of vertices of the 4D~cube on which the desired 24-tuple of quad-equations will lie. The only problem that may occur is that the biquadratics patterns and colorings of vertices of the original B\"acklund cubes are not compatible. We prove that this is not the case by just presenting the biquadratics patterns of all possible constellations (see Figures~\ref{fig:classi1}--\ref{fig:classi9} on pages~\pageref{fig:classi1}--\pageref{fig:classi9}).\par
Now, we apply M\"obius transformations on the fields of these two B\"acklund cubes such that they are in canonical form, i.e., all equations are of the forms $H_{1}^{\epsilon}$--$H_{3}^{\epsilon}$, $Q_{1}^{\epsilon}$--$Q_{3}^{\epsilon}$ or $Q_{4}$. Then the desired 24-tuple of quad-equations can be constructed by just completing the six-tuples on the remaining 3D facets with equations such that they are in canonical form with parameters induced from tetrahedron equations. Due to the results of the classification in \cite{classification}, there is precisely one way to do this. The only remaining task is now to retransform the fields of the original B\"acklund cubes in order to get the desired 24-tuple of quad-equations up to M\"obius transformations of fields not belonging to the original B\"acklund cubes.
\end{proof}
\end{theo}
Considering now the whole lattice $\Z^{2}$, we get the following corollary:
\begin{cor}[Bianchi permutability of elementary B\"acklund transformations]\label{cor}
Let $x$ be a solution of a system of quad-equations on $\Z^{2}\cong\Z^{2}\times\left\{0\right\}\times\left\{0\right\}$ generated from the equation $A\left(x,x_{1},x_{2},x_{12}\right)=0$ by the reflection procedure described in Construction~\ref{const1} and two of its B\"acklund transformations $x\overset{\lambda_{1}}{\mapsto}y^{\left(1\right)}$ and $x\overset{\lambda_{2}}{\mapsto}y^{\left(2\right)}$ (depending on one value $y^{\left(1\right)}\left(0\right)$ respectively on $y^{\left(2\right)}$ each).\par
Then due to Construction~\ref{const2} and Theorem~\ref{th:bianchi}, the two corresponding layers of B\"acklund cubes on $\Z^{2}\times\left\{0,1\right\}\times\left\{0\right\}$ and $\Z^{2}\times\left\{0\right\}\times\left\{0,1\right\}$, respectively, can be extended to a layer of Bianchi cubes on $\Z^{2}\times\left\{0,1\right\}\times\left\{0,1\right\}$ which is unique up M\"obius transformations in the fields $X$, $X_{1}$, $X_{2}$, $X_{12}$ of an equation $G\left(X,X_{1},X_{2},X_{12}\right)=0$ on an arbitrary face of $\Z^{2}\times\left\{1\right\}\times\left\{1\right\}$.\par
Furthermore, there exists a unique solution $X$ on the system of quad-equations on $\Z^{2}\times\left\{1\right\}\times\left\{1\right\}$ which is obtained from $G\left(X,X_{1},X_{2},X_{12}\right)=0$ by the procedure of reflections described in Construction~\ref{const1} which is simultaneously a B\"acklund transform of $y^{\left(1\right)}$ via a B\"acklund transformation $y^{\left(1\right)}\overset{\lambda_{2}}{\mapsto} X$ and of $y^{\left(2\right)}$ via a B\"acklund transformation $y^{\left(2\right)}\overset{\lambda_{1}}{\mapsto} X$.
\end{cor}
The Bianchi permutability can be described by the so-called \emph{Bianchi diagrams} (see Figure~\ref{fig:H1Bianchi}). The corresponding \emph{superposition principle}, i.e.\ the relation between the vertices $x_{i}$, $y_{i}^{\left(1\right)}$, $y_{i}^{\left(2\right)}$ and $X_{i}$, can also be described by quad-equations.\par
In the case of quad-equations from the ABS-list and B\"acklund transformations which can be also described by the same equations from the ABS-list it was already proven that the Bianchi permutability is a direct consequence of the 4D consistency (see~\cite{SIDE}). This is also the case for our systems, but we had to check that for every given quad-equation and for every two of its B\"acklund transformations the corresponding 3D cubes fits together as two neighboring 3D facets in a 4D cube.

\section{Bianchi permutability of auto-\texorpdfstring{B\"acklund}{Baecklund} transformations} \label{auto}
As we mentioned in Section~\ref{cubes}, our definition of 3D consistency leads to the existence of (properly generalized) auto-B\"acklund transformations. Now, we will show that our definition of 4D consistency allows for a proof of the Bianchi permutability of these auto-B\"acklund transformations.\par
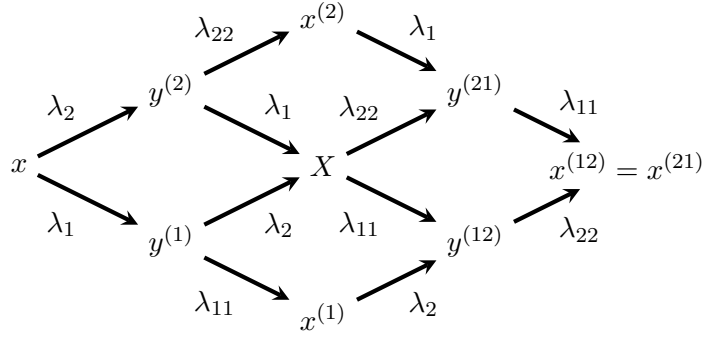
\begin{figure}[htbp]
   \centering
   \begin{tikzpicture}[auto,>=stealth]
      \node (x) at (0,0) {$x$};
      \node (y1) at (2,-1) {$y^{\left(1\right)}$};
      \node (y2) at (2,1) {$y^{\left(2\right)}$};
      \node (X) at (4,0) {$X$};
      \node (x1) at (4,-2) {$x^{\left(1\right)}$};
      \node (x2) at (4,2) {$x^{\left(2\right)}$};
      \node (y12) at (6,-1) {$y^{\left(12\right)}$};
      \node (y21) at (6,1) {$y^{\left(21\right)}$};
      \node (x12) at (8,0) {$x^{\left(12\right)}=x^{\left(21\right)}$};
      \draw [ultra thick,->] (x) to node [swap] {$\lambda_{1}$} (y1);
      \draw [ultra thick,->] (x) to node {$\lambda_{2}$} (y2);
      \draw [ultra thick,->] (y2) to node {$\lambda_{1}$} (X);
      \draw [ultra thick,->] (y1) to node [swap] {$\lambda_{2}$} (X);
      \draw [ultra thick,->] (y1) to node [swap] {$\lambda_{11}$} (x1);
      \draw [ultra thick,->] (y2) to node {$\lambda_{22}$} (x2);
      \draw [ultra thick,->] (X) to node [swap] {$\lambda_{11}$} (y12);
      \draw [ultra thick,->] (x1) to node [swap] {$\lambda_{2}$} (y12);
      \draw [ultra thick,->] (X) to node {$\lambda_{22}$} (y21);
      \draw [ultra thick,->] (x2) to node {$\lambda_{1}$} (y21);
      \draw [ultra thick,->] (y21) to node {$\lambda_{11}$} (x12);
      \draw [ultra thick,->] (y12) to node [swap] {$\lambda_{22}$} (x12);
   \end{tikzpicture}
   \caption{Bianchi permutability of auto-B\"acklund transformations}
   \label{fig:Bianchi1}
\end{figure}
Due to the generalized definition of auto-B\"acklund transformations our \emph{Bianchi diagram} is also a bit more complicated (see Figure~\ref{fig:Bianchi1}). We have the following theorem:
\begin{theo}[Bianchi permutability of auto-B\"acklund transformations]\label{th:Bianchi2}
Consider a solution $x$ of a system of quad-equations on $\Z^{2}\cong\Z^{2}\times\left\{0\right\}\times\left\{0\right\}$ generated by the reflection procedure described in Construction~\ref{const1} and two two-stage auto-B\"acklund transformations
\[
aBT_{1}:x\overset{\lambda_{1}}{\mapsto}y^{\left(1\right)}\overset{\lambda_{11}}{\mapsto}x^{\left(1\right)}
\]
and
\[
aBT_{2}:x\overset{\lambda_{2}}{\mapsto}y^{\left(2\right)}\overset{\lambda_{22}}{\mapsto}x^{\left(2\right)}
\]
(depending on two values $y^{\left(1\right)}\left(0\right)$ and $x^{\left(1\right)}\left(0\right)$ respectively on $y^{\left(2\right)}\left(0\right)$ and $x^{\left(2\right)}\left(0\right)$ each).\par
Then there exists a unique solution $x^{\left(12\right)}=x^{\left(21\right)}$ which is simultaneously and B\"acklund transform of $x^{\left(1\right)}$ via the two-stage auto-B\"acklund transformation
\[
aBT_{2}:x^{\left(1\right)}\overset{\lambda_{2}}{\mapsto}y^{\left(12\right)}\overset{\lambda_{22}}{\mapsto}x^{\left(12\right)}
\]
and of $x^{\left(2\right)}$ via the two-stage auto-B\"acklund transformation
\[
aBT_{1}:x^{\left(2\right)}\overset{\lambda_{1}}{\mapsto}y^{\left(21\right)}\overset{\lambda_{11}}{\mapsto}x^{\left(21\right)}.
\]
\begin{proof}
The B\"acklund transforms $y^{\left(1\right)}$ and $y^{\left(2\right)}$ (satisfying systems of quad-equations on $\Z^{2}\times\left\{1\right\}\times\left\{0\right\}$ and $\Z^{2}\times\left\{0\right\}\times\left\{1\right\}$, respectively) via the B\"acklund transformations $BT_{i}:x\overset{\lambda_{i}}{\mapsto}y^{\left(1\right)}$ are completely defined by the parameters $\lambda_{i}$ and values $y^{\left(i\right)}\left(0\right)$. Then due to Corollary~\ref{cor} there exists a system of quad-equations on $\Z^{2}\times\left\{1\right\}\times\left\{1\right\}$ which is obtained by an equation $G\left(X,X_{1},X_{2},X_{12}\right)=0$ (unique up to M\"obius transformations in the fields $X$, $X_{1}$, $X_{2}$ and $X_{12}$) and a unique solution $X$ of this system which is a B\"acklund transformation of $y^{\left(1\right)}$ and $y\left(2\right)$ simultaneously.\par
The B\"acklund transforms $x^{\left(1\right)}$ and $x^{\left(2\right)}$ (satisfying systems of quad-equations on $\Z^{2}\times\left\{2\right\}\times\left\{0\right\}$ and $\Z^{2}\times\left\{0\right\}\times\left\{2\right\}$, respectively) via the B\"acklund transformations $BT_{ii}:y^{\left(i\right)}\overset{\lambda_{ii}}{\mapsto}x^{\left(i\right)}$ are completely defined by the parameters $\lambda_{ii}$ and values $x^{\left(i\right)}\left(0\right)$. Then again due to Corollary~\ref{cor} there exist systems of quad-equations on $\Z^{2}\times\left\{2\right\}\times\left\{1\right\}$ and $\Z^{2}\times\left\{1\right\}\times\left\{2\right\}$ which are obtained from equations $H_{i}\left(y^{\left(ij\right)},y^{\left(ij\right)}_{1},y^{\left(ij\right)}_{2},y^{\left(ij\right)}_{12}\right)=0$ by Construction~\ref{const1} and unique solutions $y^{\left(ij\right)}$ of these systems which are B\"acklund transformations of $x^{\left(i\right)}$ and $X$ simultaneously. Here, we do not have the freedom of applying M\"obius transformations on the fields $y^{\left(ij\right)}$, $y^{\left(ij\right)}_{1}$, $y^{\left(ij\right)}_{2}$ and $y^{\left(ij\right)}_{12}$, because the corresponding Bianchi cubes have to be reflected copies of the corresponding Bianchi cubes of the previous step. Therefore, $y^{\left(ij\right)}=BT_{j}\left(x^{\left(i\right)}\right)$ holds.\par
Due to Corollary~\ref{cor} there exists a system of quad-equations on $\Z^{2}\times\left\{2\right\}\times\left\{2\right\}$ which is obtained from an equation $J\left(x^{\left(12\right)},x^{\left(12\right)}_{1},x^{\left(12\right)}_{2},x^{\left(12\right)}_{12}\right)=0$ by Construction~\ref{const1} and a unique solution $x^{\left(12\right)}=x^{\left(21\right)}$ of this system which is a B\"acklund transformation of $y^{\left(12\right)}$ and $y^{\left(21\right)}$ simultaneously. Equations on corresponding faces of $\Z^{2}\times\left\{0\right\}\times\left\{0\right\}$ and $\Z^{2}\times\left\{2\right\}\times\left\{2\right\}$ coincide, because the corresponding Bianchi cubes have to be reflected copies of the corresponding Bianchi cubes of the previous step. Therfore, $x^{\left(ij\right)}=BT_{jj}\left(y^{\left(ij\right)}\right)$ holds.\par
Finally, we want to point out that $x^{\left(12\right)}=x^{\left(21\right)}$ is independent of M\"obius transformations in the fields $X$, $X_{1}$, $X_{2}$ and $X_{12}$ in the above construction. This can be proven as follows: Consider fixed solutions $x$, $y^{\left(i\right)}$, $x^{\left(i\right)}$, $y^{\left(ij\right)}$ and $x^{\left(12\right)}=x^{\left(21\right)}$ which fit in the above setting and apply arbitrary M\"obius transformations to the fields on $\Z^{2}\times\left\{1\right\}\times\left\{1\right\}$. Then these solutions are still solutions of the set of all equations which do not depend on these fields.
\end{proof}
\end{theo}
Therefore, here the superposition principle is given by four quad-equations for every lattice point $x$ all containing the additional field $X$. However, it is sufficient to derive one of these four quad-equations, e.g.\ the one connecting $x$, $y^{\left(1\right)}$, $y^{\left(2\right)}$ and $X$, because the other equations are just reflected copies in the same manner as in the procedure of embedding quad-equations in $\Z^{4}$ (see Section~\ref{consi}). Nonetheless, we have to calculate four equations, because these equations differ for each field $x$, $x_{1}$, $x_{2}$ and $x_{12}$ of the original equation.\par
As an example, we will calculate these equations describing the super-position principle of two auto-B\"acklund transformations of the system of quad-equations on $\Z^{2}\times\left\{0\right\}\times\left\{0\right\}$ generated from equation~\eqref{eq:H1} by Construction~\ref{const1} which we already considered in the introduction. In other words, we will construct a layer of Bianchi cubes.\par
We consider the auto-B\"acklund transformation $aBT_{1}:x\overset{\lambda_{1}}{\mapsto}y^{\left(1\right)}\overset{\lambda_{11}}{\mapsto} x^{\left(1\right)}$, where the equations describing the elementary B\"acklund transformation $BT_{1}:x\overset{\lambda_{1}}{\mapsto} y^{\left(1\right)}$ were already given in \eqref{eq:H1Baeck1}. Then $y^{\left(1\right)}$ satisfies the system of quad-equations on $\Z^{2}\times\left\{1\right\}\times\left\{0\right\}$ generated from the equation \eqref{eq:H1Baeckt1}.\par
Additionally, we consider the auto-B\"acklund transformation $aBT_{2}:x\overset{\lambda_{2}}{\mapsto}y^{\left(2\right)}\overset{\lambda_{22}}{\mapsto} x^{\left(2\right)}$ where the equations describing the elementary B\"acklund transformation $BT_{2}:x\overset{\lambda_{2}}{\mapsto} y^{\left(2\right)}$ were already given in \eqref{eq:H1Baeck2}. Then $y^{\left(2\right)}$ satisfies the system of quad-equations on $\Z^{2}\times\left\{0\right\}\times\left\{1\right\}$ generated from the equation \eqref{eq:H1Baeckt2}.\par
We want to put our original equation and the equations describing the B\"acklund transformations $BT_{i}$ on a 4D cube, where the transformation $BT_{1}$ is assigned to the third direction and the transformation $BT_{2}$ is assigned to the fourth direction. Therefore, the 4D cube has a black-and-white coloring of vertices as demonstrated in Figure~\ref{fig:classi8} on page \pageref{fig:classi8}.\par
As the system of quad-equations on $\Z^{2}\times\left\{1\right\}\times\left\{1\right\}$ we choose the one generated from the equation \eqref{eq:H1Bianchi} which is unique up to M\"obius transformations in the fields $X$, $X_{1}$, $X_{2}$ and $X_{12}$. Then the transformation $y^{\left(1\right)}\overset{\lambda_{2}}{\mapsto} X$ can be described by the equations
\begin{align*}
\left(y^{\left(1\right)}-X_{2}\right)\left(y_{2}^{\left(1\right)}-X\right)-\lambda_{2}\left(1+\epsilon^{2}y_{2}^{\left(1\right)}X\right)&=0,\\
\left(y^{\left(1\right)}-y_{1}^{\left(1\right)}\right)\left(X-X_{1}\right)-\alpha\left(1+\epsilon^{2}XX_{1}\right)&=0,\\
\left(y_{1}^{\left(1\right)}-X_{12}\right)\left(y_{12}^{\left(1\right)}-X_{1}\right)-\lambda_{2}\left(1+\epsilon^{2}y_{12}^{\left(1\right)}X_{1}\right)&=0,\\
\left(y_{2}^{\left(1\right)}-y_{12}^{\left(1\right)}\right)\left(X_{2}-X_{12}\right)-\alpha\left(1+\epsilon^{2}y_{2}^{\left(1\right)}y_{12}^{\left(1\right)}\right)&=0
\end{align*}
and the transformation $y^{\left(2\right)}\overset{\lambda_{1}}{\mapsto} X$ can be described by the equations
\begin{gather*}
\left(y^{\left(2\right)}-X\right)\left(y_{2}^{\left(2\right)}-X_{2}\right)-\lambda_{1}\left(1+\epsilon^{2}y^{\left(2\right)}X\right)=0,\\
\lambda_{1}\left(y^{\left(2\right)}-y_{1}^{\left(2\right)}\right)\left(X-X_{1}\right)-\alpha\left(y^{\left(2\right)}-X\right)\left(y_{1}^{\left(2\right)}-X_{1}\right)=0,\\
\left(y_{1}^{\left(2\right)}-X_{1}\right)\left(y_{12}^{\left(2\right)}-X_{12}\right)-\lambda_{1}\left(1+\epsilon^{2}y_{1}^{\left(2\right)}X_{1}\right)=0,\\
\lambda_{1}\left(y_{2}^{\left(2\right)}-y_{12}^{\left(2\right)}\right)\left(X_{2}-X_{12}\right)-\alpha\left(y_{2}^{\left(2\right)}-X_{2}\right)\left(y_{12}^{\left(2\right)}-X_{12}\right)+\epsilon^{2}\lambda_{1}\alpha\left(\alpha-\lambda_{1}\right)=0.
\end{gather*}
Because the six-tuples of quad-equations on all 3D facets of our 4D cube have to be 3D~consistent, the equations describing the superposition principle are the following
\begin{equation}\label{eq:H1super}
\begin{aligned}
\left(x-y^{\left(1\right)}\right)\left(y^{\left(2\right)}-X\right)-\lambda_{1}\left(1+\epsilon^{2}y^{\left(2\right)}X\right)&=0,\\
\left(x_{1}-y_{1}^{\left(1\right)}\right)\left(y_{1}^{\left(2\right)}-X_{1}\right)-\lambda_{1}\left(1+\epsilon^{2}y_{1}^{\left(2\right)}X_{1}\right)&=0,\\
\left(x_{2}-y_{2}^{\left(1\right)}\right)\left(y_{2}^{\left(2\right)}-X_{2}\right)-\lambda_{1}\left(1+\epsilon^{2}x_{2}y_{2}^{\left(1\right)}\right)&=0,\\
\left(x_{12}-y_{12}^{\left(1\right)}\right)\left(y_{12}^{\left(2\right)}-X_{12}\right)-\lambda_{1}\left(1+\epsilon^{2}x_{12}y_{12}^{\left(1\right)}\right)&=0.
\end{aligned}
\end{equation}
The resulting 24-tuple is 4D consistent, because all six-tuples on the 3D facets of the corresponding 4D cube are 3D consistent. The complete layer of Bianchi cubes can be obtained by the reflection procedure described in Construction~\ref{const4}.

\section{Conclusion}
By proving the Bianchi permutability of auto-B\"acklund transformations, we presented the maybe last missing argument for using 3D consistency as a definition of integrability of quad-equations even in the generalized version from \cite{ABS2,classification}. This was only possible by the consideration of 4D consistent systems of quad-equations with all the difficulties of embedding in the 4D lattice and the important object of super-consistent eight-tuples on decorated 3D cubes, i.e., the remarkable fact that the two quad-equations describing the tetrahedron property of a B\"acklund cube can be considered as equations on equal footing with the equations of the B\"acklund cube.

\section*{Acknowledgments}
The author is supported by DFG (Deutsche Forschungsgemeinschaft) in the frame of Sonderforschungsbereich/Transregio 109 “Discretization in Geometry and Dynamics” and is indebted to Yuri~B. Suris for his continued guidance.
\appendix
\section[Biquadratics patterns]{Biquadratics patterns of the 4D cubes for all possible pairs of B\"acklund transformations}
Here, we present by just giving the black-and-white coloring for clarity. However, one can easily see the biquadratics pattern from these figures, because edges carrying non-degenerate biquadratics always connect vertices of the same color and edges carrying degenerate biquadratics always connect vertices of different colors.
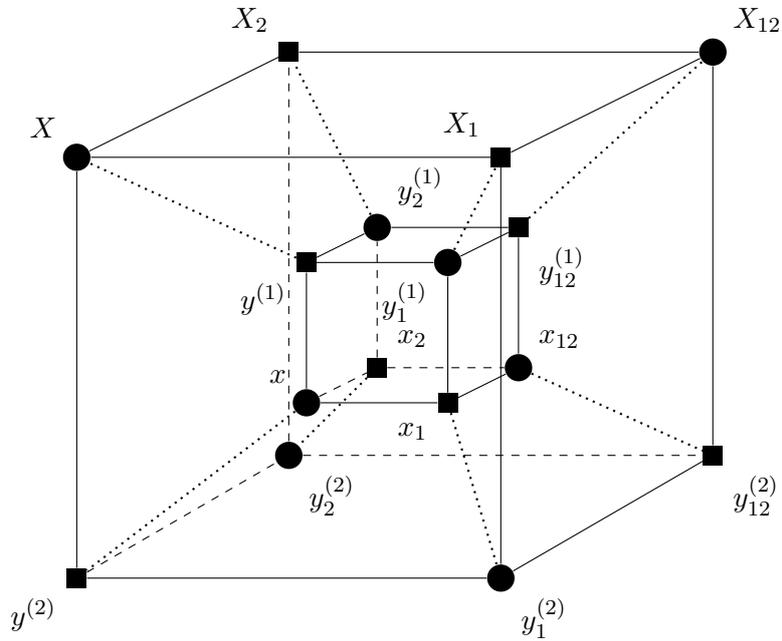
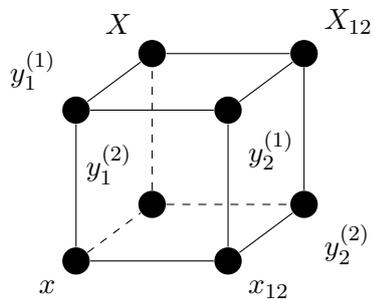
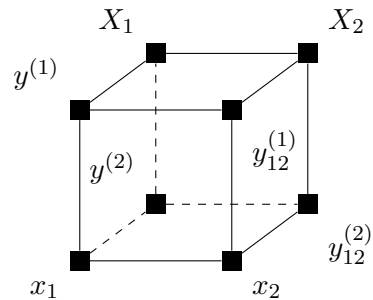
\begin{figure}[htbp]
   \centering
   \subfloat[Biquadratics pattern for the 4D cube]{
   \begin{tikzpicture}[scale=0.93]
      \node (x) at (0,0) [circle,fill,label=135:$x$] {};
      \node (x1) at (2,0) [fill,label=-135:$x_{1}$] {};
      \node (x2) at (1,0.5) [fill,label=45:$x_{2}$] {};
      \node (x3) at (0,2) [fill,label=-135:$y^{\left(1\right)}$] {};
      \node (x4) at (-3.25,-2.5) [fill,label=-135:$y^{\left(2\right)}$] {};
      \node (x12) at (3,0.5) [circle,fill,label=45:$x_{12}$] {};
      \node (x13) at (2,2) [circle,fill,label=-135:$y_{1}^{\left(1\right)}$] {};
      \node (x14) at (2.75,-2.5) [circle,fill,label=-45:$y_{1}^{\left(2\right)}$] {};
      \node (x23) at (1,2.5) [circle,fill,label=45:$y_{2}^{\left(1\right)}$] {};
      \node (x24) at (-0.25,-0.75) [circle,fill,label=-45:$y_{2}^{\left(2\right)}$] {};
      \node (x34) at (-3.25,3.5) [circle,fill,label=135:$X$] {};
      \node (x123) at (3,2.5) [fill,label=-45:$y_{12}^{\left(1\right)}$] {};
      \node (x124) at (5.75,-0.75) [fill,label=-45:$y_{12}^{\left(2\right)}$] {};
      \node (x134) at (2.75,3.5) [fill,label=135:$X_{1}$] {};
      \node (x234) at (-0.25,5) [fill,label=135:$X_{2}$] {};
      \node (x1234) at (5.75,5) [circle,fill,label=45:$X_{12}$] {};
      \draw (x) to (x1) to (x12) to (x123) to (x23) to (x3) to (x);
      \draw (x4) to (x14) to (x124) to (x1234) to (x234) to (x34) to (x4);
      \draw (x1) to (x13) to (x3);
      \draw (x14) to (x134) to (x34);
      \draw (x13) to (x123);
      \draw (x134) to (x1234);
      \draw [dashed] (x) to (x2) to (x12);
      \draw [dashed] (x4) to (x24) to (x124);
      \draw [dashed] (x2) to (x23);
      \draw [dashed] (x24) to (x234);
      \draw [dotted,thick] (x) to (x4);
      \draw [dotted,thick] (x1) to (x14);
      \draw [dotted,thick] (x2) to (x24);
      \draw [dotted,thick] (x3) to (x34);
      \draw [dotted,thick] (x12) to (x124);
      \draw [dotted,thick] (x13) to (x134);
      \draw [dotted,thick] (x23) to (x234);
      \draw [dotted,thick] (x123) to (x1234);
   \end{tikzpicture}
   }\\
   \subfloat[Biquadratics pattern of the 3D cube for $\T$]{
      \begin{tikzpicture}
      \node (x) at (0,0) [circle,fill,label=-135:$x$] {};
      \node (x1) at (2,0) [circle,fill,label=-45:$x_{12}$] {};
      \node (x2) at (1,0.75) [circle,fill,label=135:$y_{1}^{\left(2\right)}$] {};
      \node (x12) at (3,0.75) [circle,fill,label=-45:$y_{2}^{\left(2\right)}$] {};
      \node (x3) at (0,2) [circle,fill,label=135:$y_{1}^{\left(1\right)}$] {};
      \node (x13) at (2,2) [circle,fill,label=-45:$y_{2}^{\left(1\right)}$] {};
      \node (x23) at (1,2.75) [circle,fill,label=135:$X$] {};
      \node (x123) at (3,2.75) [circle,fill,label=45:$X_{12}$] {};
      \draw (x) to (x1) to (x12) to (x123) to (x23) to (x3) to (x);
      \draw (x1) to (x13) to (x3);
      \draw (x13) to (x123);
      \draw [dashed] (x) to (x2) to (x12);
      \draw [dashed] (x2) to (x23);
   \end{tikzpicture}
   }\qquad
   \subfloat[Biquadratics pattern of the 3D cube for $\bar{\T}$]{
   \begin{tikzpicture}
      \node (x) at (0,0) [fill,label=-135:$x_{1}$] {};
      \node (x1) at (2,0) [fill,label=-45:$x_{2}$] {};
      \node (x2) at (1,0.75) [fill,label=135:$y^{\left(2\right)}$] {};
      \node (x12) at (3,0.75) [fill,label=-45:$y_{12}^{\left(2\right)}$] {};
      \node (x3) at (0,2) [fill,label=135:$y^{\left(1\right)}$] {};
      \node (x13) at (2,2) [fill,label=-45:$y_{12}^{\left(1\right)}$] {};
      \node (x23) at (1,2.75) [fill,label=135:$X_{1}$] {};
      \node (x123) at (3,2.75) [fill,label=45:$X_{2}$] {};
      \draw (x) to (x1) to (x12) to (x123) to (x23) to (x3) to (x);
      \draw (x1) to (x13) to (x3);
      \draw (x13) to (x123);
      \draw [dashed] (x) to (x2) to (x12);
      \draw [dashed] (x2) to (x23);
   \end{tikzpicture}
   }
   \caption{Two B\"acklund transforms $y^{\left(1\right)}$ and $y^{\left(2\right)}$ of the solution $x$ of a type~Q equation, where both B\"acklund transformations can be described by type~Q equations.}
   \label{fig:classi1}
\end{figure}
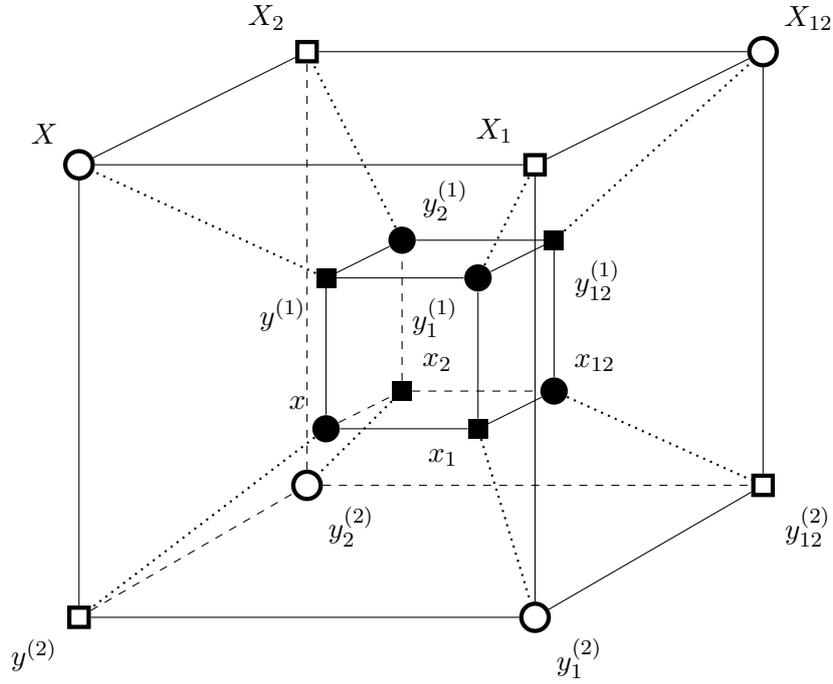
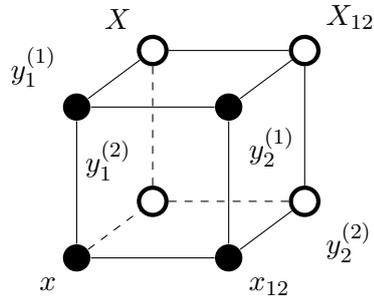
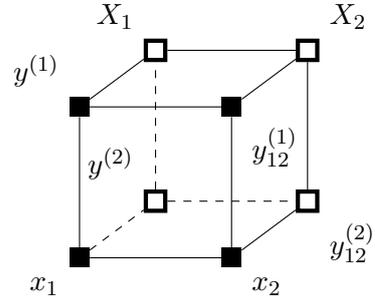
\begin{figure}[htbp]
   \centering
   \subfloat[Biquadratics pattern for the 4D cube]{
   \begin{tikzpicture}
      \node (x) at (0,0) [circle,fill,label=135:$x$] {};
      \node (x1) at (2,0) [fill,label=-135:$x_{1}$] {};
      \node (x2) at (1,0.5) [fill,label=45:$x_{2}$] {};
      \node (x3) at (0,2) [fill,label=-135:$y^{\left(1\right)}$] {};
      \node (x4) at (-3.25,-2.5) [draw,ultra thick,label=-135:$y^{\left(2\right)}$] {};
      \node (x12) at (3,0.5) [circle,fill,label=45:$x_{12}$] {};
      \node (x13) at (2,2) [circle,fill,label=-135:$y_{1}^{\left(1\right)}$] {};
      \node (x14) at (2.75,-2.5) [circle,draw,ultra thick,label=-45:$y_{1}^{\left(2\right)}$] {};
      \node (x23) at (1,2.5) [circle,fill,label=45:$y_{2}^{\left(1\right)}$] {};
      \node (x24) at (-0.25,-0.75) [circle,draw,ultra thick,label=-45:$y_{2}^{\left(2\right)}$] {};
      \node (x34) at (-3.25,3.5) [circle,draw,ultra thick,label=135:$X$] {};
      \node (x123) at (3,2.5) [fill,label=-45:$y_{12}^{\left(1\right)}$] {};
      \node (x124) at (5.75,-0.75) [draw,ultra thick,label=-45:$y_{12}^{\left(2\right)}$] {};
      \node (x134) at (2.75,3.5) [draw,ultra thick,label=135:$X_{1}$] {};
      \node (x234) at (-0.25,5) [draw,ultra thick,label=135:$X_{2}$] {};
      \node (x1234) at (5.75,5) [circle,draw,ultra thick,label=45:$X_{12}$] {};
      \draw (x) to (x1) to (x12) to (x123) to (x23) to (x3) to (x);
      \draw (x4) to (x14) to (x124) to (x1234) to (x234) to (x34) to (x4);
      \draw (x1) to (x13) to (x3);
      \draw (x14) to (x134) to (x34);
      \draw (x13) to (x123);
      \draw (x134) to (x1234);
      \draw [dashed] (x) to (x2) to (x12);
      \draw [dashed] (x4) to (x24) to (x124);
      \draw [dashed] (x2) to (x23);
      \draw [dashed] (x24) to (x234);
      \draw [dotted,thick] (x) to (x4);
      \draw [dotted,thick] (x1) to (x14);
      \draw [dotted,thick] (x2) to (x24);
      \draw [dotted,thick] (x3) to (x34);
      \draw [dotted,thick] (x12) to (x124);
      \draw [dotted,thick] (x13) to (x134);
      \draw [dotted,thick] (x23) to (x234);
      \draw [dotted,thick] (x123) to (x1234);
   \end{tikzpicture}
   }\\
   \subfloat[Biquadratics pattern of the 3D cube for $\T$]{
      \begin{tikzpicture}
      \node (x) at (0,0) [circle,fill,label=-135:$x$] {};
      \node (x1) at (2,0) [circle,fill,label=-45:$x_{12}$] {};
      \node (x2) at (1,0.75) [circle,draw,ultra thick,label=135:$y_{1}^{\left(2\right)}$] {};
      \node (x12) at (3,0.75) [circle,draw,ultra thick,label=-45:$y_{2}^{\left(2\right)}$] {};
      \node (x3) at (0,2) [circle,fill,label=135:$y_{1}^{\left(1\right)}$] {};
      \node (x13) at (2,2) [circle,fill,label=-45:$y_{2}^{\left(1\right)}$] {};
      \node (x23) at (1,2.75) [circle,draw,ultra thick,label=135:$X$] {};
      \node (x123) at (3,2.75) [circle,draw,ultra thick,label=45:$X_{12}$] {};
      \draw (x) to (x1) to (x12) to (x123) to (x23) to (x3) to (x);
      \draw (x1) to (x13) to (x3);
      \draw (x13) to (x123);
      \draw [dashed] (x) to (x2) to (x12);
      \draw [dashed] (x2) to (x23);
   \end{tikzpicture}
   }\qquad
   \subfloat[Biquadratics pattern of the 3D cube for $\bar{\T}$]{
   \begin{tikzpicture}
      \node (x) at (0,0) [fill,label=-135:$x_{1}$] {};
      \node (x1) at (2,0) [fill,label=-45:$x_{2}$] {};
      \node (x2) at (1,0.75) [draw,ultra thick,label=135:$y^{\left(2\right)}$] {};
      \node (x12) at (3,0.75) [draw,ultra thick,label=-45:$y_{12}^{\left(2\right)}$] {};
      \node (x3) at (0,2) [fill,label=135:$y^{\left(1\right)}$] {};
      \node (x13) at (2,2) [fill,label=-45:$y_{12}^{\left(1\right)}$] {};
      \node (x23) at (1,2.75) [draw,ultra thick,label=135:$X_{1}$] {};
      \node (x123) at (3,2.75) [draw,ultra thick,label=45:$X_{2}$] {};
      \draw (x) to (x1) to (x12) to (x123) to (x23) to (x3) to (x);
      \draw (x1) to (x13) to (x3);
      \draw (x13) to (x123);
      \draw [dashed] (x) to (x2) to (x12);
      \draw [dashed] (x2) to (x23);
   \end{tikzpicture}
   }
   \caption{Two B\"acklund transforms $y^{\left(1\right)}$ and $y^{\left(2\right)}$ of the solution $x$ of a type~Q equation, where one B\"acklund transformation can be described by type~Q equations and the other B\"acklund transformation can be described by type~\Hvier\ equations in the trapezoidal version.}
   \label{fig:classi2}
\end{figure}
\begin{figure}[htbp]
   \centering
   \subfloat[Biquadratics pattern for the 4D cube]{
   \begin{tikzpicture}
      \node (x) at (0,0) [circle,fill,label=135:$x$] {};
      \node (x1) at (2,0) [fill,label=-135:$x_{1}$] {};
      \node (x2) at (1,0.5) [fill,label=45:$x_{2}$] {};
      \node (x3) at (0,2) [draw,ultra thick,label=-135:$y^{\left(1\right)}$] {};
      \node (x4) at (-3.25,-2.5) [draw,ultra thick,label=-135:$y^{\left(2\right)}$] {};
      \node (x12) at (3,0.5) [circle,fill,label=45:$x_{12}$] {};
      \node (x13) at (2,2) [circle,draw,ultra thick,label=-135:$y_{1}^{\left(1\right)}$] {};
      \node (x14) at (2.75,-2.5) [circle,draw,ultra thick,label=-45:$y_{1}^{\left(2\right)}$] {};
      \node (x23) at (1,2.5) [circle,draw,ultra thick,label=45:$y_{2}^{\left(1\right)}$] {};
      \node (x24) at (-0.25,-0.75) [circle,draw,ultra thick,label=-45:$y_{2}^{\left(2\right)}$] {};
      \node (x34) at (-3.25,3.5) [circle,fill,label=135:$X$] {};
      \node (x123) at (3,2.5) [draw,ultra thick,label=-45:$y_{12}^{\left(1\right)}$] {};
      \node (x124) at (5.75,-0.75) [draw,ultra thick,label=-45:$y_{12}^{\left(2\right)}$] {};
      \node (x134) at (2.75,3.5) [fill,label=135:$X_{1}$] {};
      \node (x234) at (-0.25,5) [fill,label=135:$X_{2}$] {};
      \node (x1234) at (5.75,5) [circle,fill,label=45:$X_{12}$] {};
      \draw (x) to (x1) to (x12) to (x123) to (x23) to (x3) to (x);
      \draw (x4) to (x14) to (x124) to (x1234) to (x234) to (x34) to (x4);
      \draw (x1) to (x13) to (x3);
      \draw (x14) to (x134) to (x34);
      \draw (x13) to (x123);
      \draw (x134) to (x1234);
      \draw [dashed] (x) to (x2) to (x12);
      \draw [dashed] (x4) to (x24) to (x124);
      \draw [dashed] (x2) to (x23);
      \draw [dashed] (x24) to (x234);
      \draw [dotted,thick] (x) to (x4);
      \draw [dotted,thick] (x1) to (x14);
      \draw [dotted,thick] (x2) to (x24);
      \draw [dotted,thick] (x3) to (x34);
      \draw [dotted,thick] (x12) to (x124);
      \draw [dotted,thick] (x13) to (x134);
      \draw [dotted,thick] (x23) to (x234);
      \draw [dotted,thick] (x123) to (x1234);
   \end{tikzpicture}
   }\\
   \subfloat[Biquadratics pattern of the 3D cube for $\T$]{
      \begin{tikzpicture}
      \node (x) at (0,0) [circle,fill,label=-135:$x$] {};
      \node (x1) at (2,0) [circle,fill,label=-45:$x_{12}$] {};
      \node (x2) at (1,0.75) [circle,draw,ultra thick,label=135:$y_{1}^{\left(2\right)}$] {};
      \node (x12) at (3,0.75) [circle,draw,ultra thick,label=-45:$y_{2}^{\left(2\right)}$] {};
      \node (x3) at (0,2) [circle,draw,ultra thick,label=135:$y_{1}^{\left(1\right)}$] {};
      \node (x13) at (2,2) [circle,draw,ultra thick,label=-45:$y_{2}^{\left(1\right)}$] {};
      \node (x23) at (1,2.75) [circle,fill,label=135:$X$] {};
      \node (x123) at (3,2.75) [circle,fill,label=45:$X_{12}$] {};
      \draw (x) to (x1) to (x12) to (x123) to (x23) to (x3) to (x);
      \draw (x1) to (x13) to (x3);
      \draw (x13) to (x123);
      \draw [dashed] (x) to (x2) to (x12);
      \draw [dashed] (x2) to (x23);
   \end{tikzpicture}
   }\qquad
   \subfloat[Biquadratics pattern of the 3D cube for $\bar{\T}$]{
   \begin{tikzpicture}
      \node (x) at (0,0) [fill,label=-135:$x_{1}$] {};
      \node (x1) at (2,0) [fill,label=-45:$x_{2}$] {};
      \node (x2) at (1,0.75) [draw,ultra thick,label=135:$y^{\left(2\right)}$] {};
      \node (x12) at (3,0.75) [draw,ultra thick,label=-45:$y_{12}^{\left(2\right)}$] {};
      \node (x3) at (0,2) [draw,ultra thick,label=135:$y^{\left(1\right)}$] {};
      \node (x13) at (2,2) [draw,ultra thick,label=-45:$y_{12}^{\left(1\right)}$] {};
      \node (x23) at (1,2.75) [fill,label=135:$X_{1}$] {};
      \node (x123) at (3,2.75) [fill,label=45:$X_{2}$] {};
      \draw (x) to (x1) to (x12) to (x123) to (x23) to (x3) to (x);
      \draw (x1) to (x13) to (x3);
      \draw (x13) to (x123);
      \draw [dashed] (x) to (x2) to (x12);
      \draw [dashed] (x2) to (x23);
   \end{tikzpicture}
   }
   \caption{Two B\"acklund transforms $y^{\left(1\right)}$ and $y^{\left(2\right)}$ of the solution $x$ of a type~Q equation, where both B\"acklund transformations can be described by type~\Hvier\ equations in the trapezoidal version.}
   \label{fig:classi3}
\end{figure}
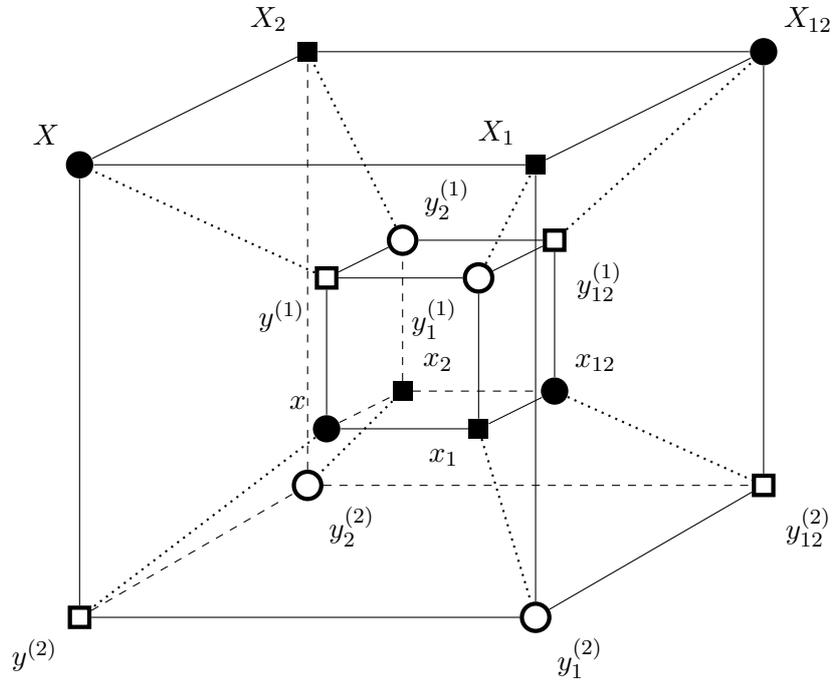
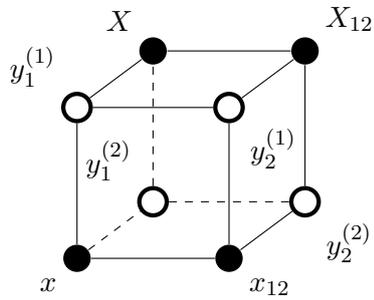
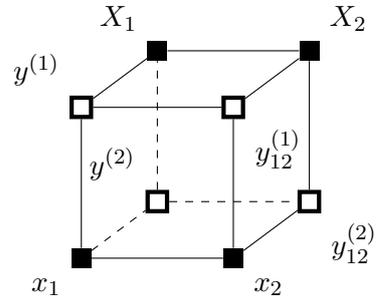
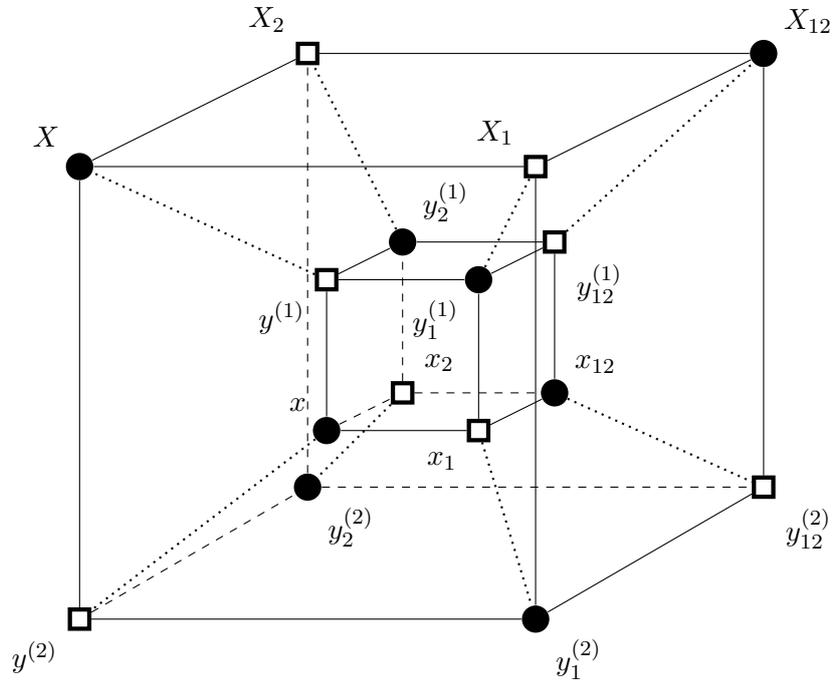
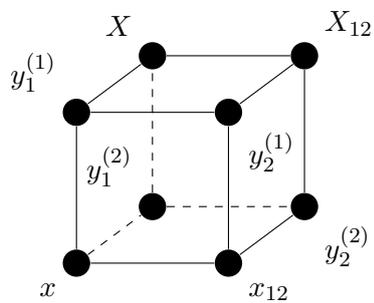
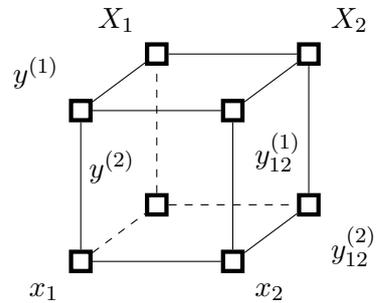
\begin{figure}[htbp]
   \centering
   \subfloat[Biquadratics pattern for the 4D cube]{
   \begin{tikzpicture}
      \node (x) at (0,0) [circle,fill,label=135:$x$] {};
      \node (x1) at (2,0) [draw,ultra thick,label=-135:$x_{1}$] {};
      \node (x2) at (1,0.5) [draw,ultra thick,label=45:$x_{2}$] {};
      \node (x3) at (0,2) [draw,ultra thick,label=-135:$y^{\left(1\right)}$] {};
      \node (x4) at (-3.25,-2.5) [draw,ultra thick,label=-135:$y^{\left(2\right)}$] {};
      \node (x12) at (3,0.5) [circle,fill,label=45:$x_{12}$] {};
      \node (x13) at (2,2) [circle,fill,label=-135:$y_{1}^{\left(1\right)}$] {};
      \node (x14) at (2.75,-2.5) [circle,fill,label=-45:$y_{1}^{\left(2\right)}$] {};
      \node (x23) at (1,2.5) [circle,fill,label=45:$y_{2}^{\left(1\right)}$] {};
      \node (x24) at (-0.25,-0.75) [circle,fill,label=-45:$y_{2}^{\left(2\right)}$] {};
      \node (x34) at (-3.25,3.5) [circle,fill,label=135:$X$] {};
      \node (x123) at (3,2.5) [draw,ultra thick,label=-45:$y_{12}^{\left(1\right)}$] {};
      \node (x124) at (5.75,-0.75) [draw,ultra thick,label=-45:$y_{12}^{\left(2\right)}$] {};
      \node (x134) at (2.75,3.5) [draw,ultra thick,label=135:$X_{1}$] {};
      \node (x234) at (-0.25,5) [draw,ultra thick,label=135:$X_{2}$] {};
      \node (x1234) at (5.75,5) [circle,fill,label=45:$X_{12}$] {};
      \draw (x) to (x1) to (x12) to (x123) to (x23) to (x3) to (x);
      \draw (x4) to (x14) to (x124) to (x1234) to (x234) to (x34) to (x4);
      \draw (x1) to (x13) to (x3);
      \draw (x14) to (x134) to (x34);
      \draw (x13) to (x123);
      \draw (x134) to (x1234);
      \draw [dashed] (x) to (x2) to (x12);
      \draw [dashed] (x4) to (x24) to (x124);
      \draw [dashed] (x2) to (x23);
      \draw [dashed] (x24) to (x234);
      \draw [dotted,thick] (x) to (x4);
      \draw [dotted,thick] (x1) to (x14);
      \draw [dotted,thick] (x2) to (x24);
      \draw [dotted,thick] (x3) to (x34);
      \draw [dotted,thick] (x12) to (x124);
      \draw [dotted,thick] (x13) to (x134);
      \draw [dotted,thick] (x23) to (x234);
      \draw [dotted,thick] (x123) to (x1234);
   \end{tikzpicture}
   }\\
   \subfloat[Biquadratics pattern of the 3D cube for $\T$]{
      \begin{tikzpicture}
      \node (x) at (0,0) [circle,fill,label=-135:$x$] {};
      \node (x1) at (2,0) [circle,fill,label=-45:$x_{12}$] {};
      \node (x2) at (1,0.75) [circle,fill,label=135:$y_{1}^{\left(2\right)}$] {};
      \node (x12) at (3,0.75) [circle,fill,label=-45:$y_{2}^{\left(2\right)}$] {};
      \node (x3) at (0,2) [circle,fill,label=135:$y_{1}^{\left(1\right)}$] {};
      \node (x13) at (2,2) [circle,fill,label=-45:$y_{2}^{\left(1\right)}$] {};
      \node (x23) at (1,2.75) [circle,fill,label=135:$X$] {};
      \node (x123) at (3,2.75) [circle,fill,label=45:$X_{12}$] {};
      \draw (x) to (x1) to (x12) to (x123) to (x23) to (x3) to (x);
      \draw (x1) to (x13) to (x3);
      \draw (x13) to (x123);
      \draw [dashed] (x) to (x2) to (x12);
      \draw [dashed] (x2) to (x23);
   \end{tikzpicture}
   }\qquad
   \subfloat[Biquadratics pattern of the 3D cube for $\bar{\T}$]{
   \begin{tikzpicture}
      \node (x) at (0,0) [draw,ultra thick,label=-135:$x_{1}$] {};
      \node (x1) at (2,0) [draw,ultra thick,label=-45:$x_{2}$] {};
      \node (x2) at (1,0.75) [draw,ultra thick,label=135:$y^{\left(2\right)}$] {};
      \node (x12) at (3,0.75) [draw,ultra thick,label=-45:$y_{12}^{\left(2\right)}$] {};
      \node (x3) at (0,2) [draw,ultra thick,label=135:$y^{\left(1\right)}$] {};
      \node (x13) at (2,2) [draw,ultra thick,label=-45:$y_{12}^{\left(1\right)}$] {};
      \node (x23) at (1,2.75) [draw,ultra thick,label=135:$X_{1}$] {};
      \node (x123) at (3,2.75) [draw,ultra thick,label=45:$X_{2}$] {};
      \draw (x) to (x1) to (x12) to (x123) to (x23) to (x3) to (x);
      \draw (x1) to (x13) to (x3);
      \draw (x13) to (x123);
      \draw [dashed] (x) to (x2) to (x12);
      \draw [dashed] (x2) to (x23);
   \end{tikzpicture}
   }
   \caption{Two B\"acklund transforms $y^{\left(1\right)}$ and $y^{\left(2\right)}$ of the solution $x$ of a type~\Hvier\ equation in the rhombic version, where both B\"acklund transformations can be described by type~\Hvier\ equations in the rhombic version.}
   \label{fig:classi4}
\end{figure}
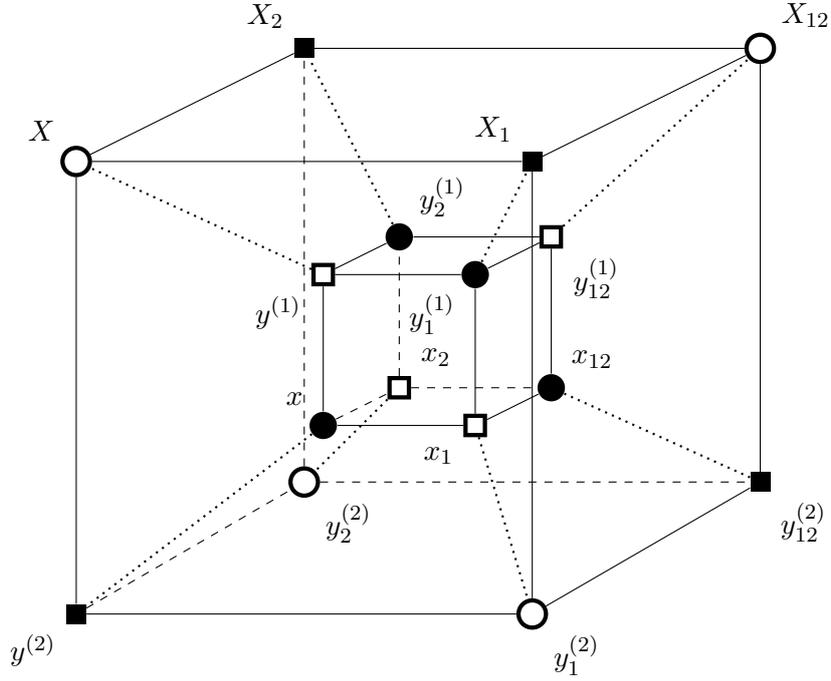
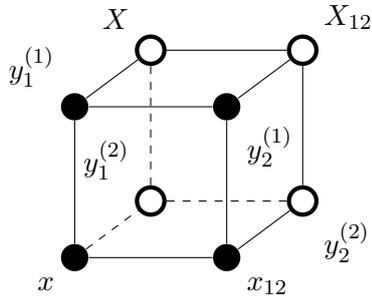
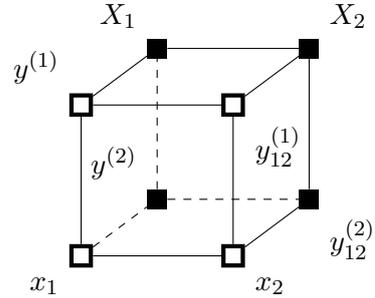
\begin{figure}[htbp]
   \centering
   \subfloat[Biquadratics pattern for the 4D cube]{
   \begin{tikzpicture}
      \node (x) at (0,0) [circle,fill,label=135:$x$] {};
      \node (x1) at (2,0) [draw,ultra thick,label=-135:$x_{1}$] {};
      \node (x2) at (1,0.5) [draw,ultra thick,label=45:$x_{2}$] {};
      \node (x3) at (0,2) [draw,ultra thick,label=-135:$y^{\left(1\right)}$] {};
      \node (x4) at (-3.25,-2.5) [fill,label=-135:$y^{\left(2\right)}$] {};
      \node (x12) at (3,0.5) [circle,fill,label=45:$x_{12}$] {};
      \node (x13) at (2,2) [circle,fill,label=-135:$y_{1}^{\left(1\right)}$] {};
      \node (x14) at (2.75,-2.5) [circle,draw,ultra thick,label=-45:$y_{1}^{\left(2\right)}$] {};
      \node (x23) at (1,2.5) [circle,fill,label=45:$y_{2}^{\left(1\right)}$] {};
      \node (x24) at (-0.25,-0.75) [circle,draw,ultra thick,label=-45:$y_{2}^{\left(2\right)}$] {};
      \node (x34) at (-3.25,3.5) [circle,draw,ultra thick,label=135:$X$] {};
      \node (x123) at (3,2.5) [draw,ultra thick,label=-45:$y_{12}^{\left(1\right)}$] {};
      \node (x124) at (5.75,-0.75) [fill,label=-45:$y_{12}^{\left(2\right)}$] {};
      \node (x134) at (2.75,3.5) [fill,label=135:$X_{1}$] {};
      \node (x234) at (-0.25,5) [fill,label=135:$X_{2}$] {};
      \node (x1234) at (5.75,5) [circle,draw,ultra thick,label=45:$X_{12}$] {};
      \draw (x) to (x1) to (x12) to (x123) to (x23) to (x3) to (x);
      \draw (x4) to (x14) to (x124) to (x1234) to (x234) to (x34) to (x4);
      \draw (x1) to (x13) to (x3);
      \draw (x14) to (x134) to (x34);
      \draw (x13) to (x123);
      \draw (x134) to (x1234);
      \draw [dashed] (x) to (x2) to (x12);
      \draw [dashed] (x4) to (x24) to (x124);
      \draw [dashed] (x2) to (x23);
      \draw [dashed] (x24) to (x234);
      \draw [dotted,thick] (x) to (x4);
      \draw [dotted,thick] (x1) to (x14);
      \draw [dotted,thick] (x2) to (x24);
      \draw [dotted,thick] (x3) to (x34);
      \draw [dotted,thick] (x12) to (x124);
      \draw [dotted,thick] (x13) to (x134);
      \draw [dotted,thick] (x23) to (x234);
      \draw [dotted,thick] (x123) to (x1234);
   \end{tikzpicture}
   }\\
   \subfloat[Biquadratics pattern of the 3D cube for $\T$]{
      \begin{tikzpicture}
      \node (x) at (0,0) [circle,fill,label=-135:$x$] {};
      \node (x1) at (2,0) [circle,fill,label=-45:$x_{12}$] {};
      \node (x2) at (1,0.75) [circle,draw,ultra thick,label=135:$y_{1}^{\left(2\right)}$] {};
      \node (x12) at (3,0.75) [circle,draw,ultra thick,label=-45:$y_{2}^{\left(2\right)}$] {};
      \node (x3) at (0,2) [circle,fill,label=135:$y_{1}^{\left(1\right)}$] {};
      \node (x13) at (2,2) [circle,fill,label=-45:$y_{2}^{\left(1\right)}$] {};
      \node (x23) at (1,2.75) [circle,draw,ultra thick,label=135:$X$] {};
      \node (x123) at (3,2.75) [circle,draw,ultra thick,label=45:$X_{12}$] {};
      \draw (x) to (x1) to (x12) to (x123) to (x23) to (x3) to (x);
      \draw (x1) to (x13) to (x3);
      \draw (x13) to (x123);
      \draw [dashed] (x) to (x2) to (x12);
      \draw [dashed] (x2) to (x23);
   \end{tikzpicture}
   }\qquad
   \subfloat[Biquadratics pattern of the 3D cube for $\bar{\T}$]{
   \begin{tikzpicture}
      \node (x) at (0,0) [draw,ultra thick,label=-135:$x_{1}$] {};
      \node (x1) at (2,0) [draw,ultra thick,label=-45:$x_{2}$] {};
      \node (x2) at (1,0.75) [fill,label=135:$y^{\left(2\right)}$] {};
      \node (x12) at (3,0.75) [fill,label=-45:$y_{12}^{\left(2\right)}$] {};
      \node (x3) at (0,2) [draw,ultra thick,label=135:$y^{\left(1\right)}$] {};
      \node (x13) at (2,2) [draw,ultra thick,label=-45:$y_{12}^{\left(1\right)}$] {};
      \node (x23) at (1,2.75) [fill,label=135:$X_{1}$] {};
      \node (x123) at (3,2.75) [fill,label=45:$X_{2}$] {};
      \draw (x) to (x1) to (x12) to (x123) to (x23) to (x3) to (x);
      \draw (x1) to (x13) to (x3);
      \draw (x13) to (x123);
      \draw [dashed] (x) to (x2) to (x12);
      \draw [dashed] (x2) to (x23);
   \end{tikzpicture}
   }
   \caption{Two B\"acklund transforms $y^{\left(1\right)}$ and $y^{\left(2\right)}$ of the solution $x$ of a type~\Hvier\ equation in the rhombic version, where one B\"acklund transformation can be described by type~\Hvier\ equations in the rhombic version and the other B\"acklund transformation can be described by type~\Hvier\ equations in the trapezoidal version.}
   \label{fig:classi5}
\end{figure}
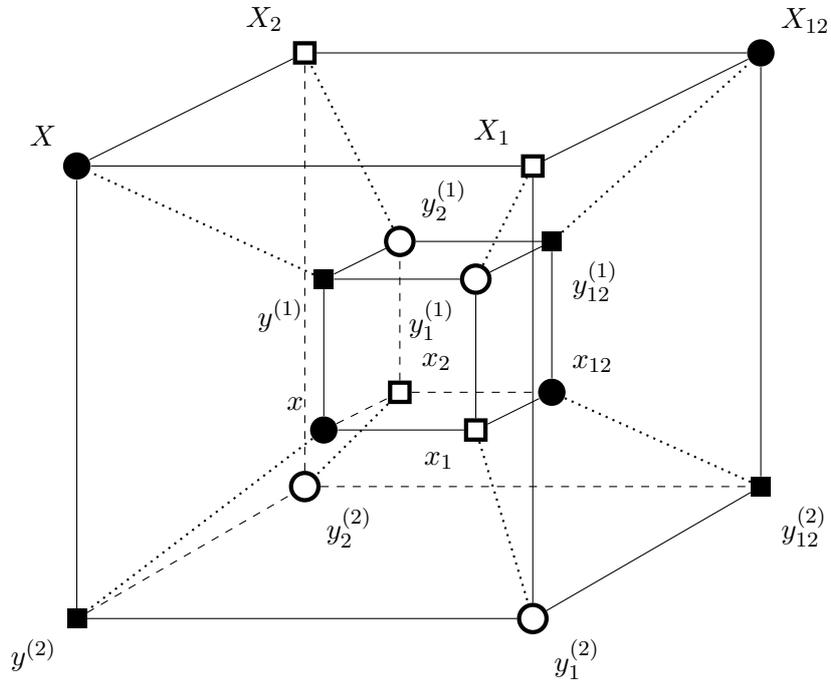
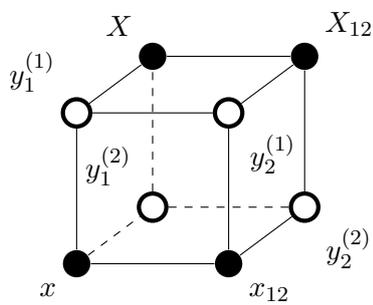
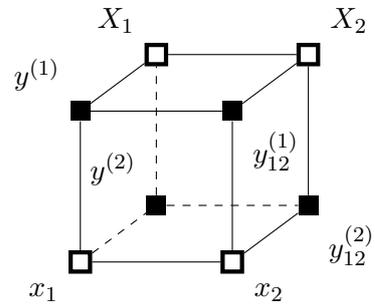
\begin{figure}[htbp]
   \centering
   \subfloat[Biquadratics pattern for the 4D cube]{
   \begin{tikzpicture}
      \node (x) at (0,0) [circle,fill,label=135:$x$] {};
      \node (x1) at (2,0) [draw,ultra thick,label=-135:$x_{1}$] {};
      \node (x2) at (1,0.5) [draw,ultra thick,label=45:$x_{2}$] {};
      \node (x3) at (0,2) [fill,label=-135:$y^{\left(1\right)}$] {};
      \node (x4) at (-3.25,-2.5) [fill,label=-135:$y^{\left(2\right)}$] {};
      \node (x12) at (3,0.5) [circle,fill,label=45:$x_{12}$] {};
      \node (x13) at (2,2) [circle,draw,ultra thick,label=-135:$y_{1}^{\left(1\right)}$] {};
      \node (x14) at (2.75,-2.5) [circle,draw,ultra thick,label=-45:$y_{1}^{\left(2\right)}$] {};
      \node (x23) at (1,2.5) [circle,draw,ultra thick,label=45:$y_{2}^{\left(1\right)}$] {};
      \node (x24) at (-0.25,-0.75) [circle,draw,ultra thick,label=-45:$y_{2}^{\left(2\right)}$] {};
      \node (x34) at (-3.25,3.5) [circle,fill,label=135:$X$] {};
      \node (x123) at (3,2.5) [fill,label=-45:$y_{12}^{\left(1\right)}$] {};
      \node (x124) at (5.75,-0.75) [fill,label=-45:$y_{12}^{\left(2\right)}$] {};
      \node (x134) at (2.75,3.5) [draw,ultra thick,label=135:$X_{1}$] {};
      \node (x234) at (-0.25,5) [draw,ultra thick,label=135:$X_{2}$] {};
      \node (x1234) at (5.75,5) [circle,fill,label=45:$X_{12}$] {};
      \draw (x) to (x1) to (x12) to (x123) to (x23) to (x3) to (x);
      \draw (x4) to (x14) to (x124) to (x1234) to (x234) to (x34) to (x4);
      \draw (x1) to (x13) to (x3);
      \draw (x14) to (x134) to (x34);
      \draw (x13) to (x123);
      \draw (x134) to (x1234);
      \draw [dashed] (x) to (x2) to (x12);
      \draw [dashed] (x4) to (x24) to (x124);
      \draw [dashed] (x2) to (x23);
      \draw [dashed] (x24) to (x234);
      \draw [dotted,thick] (x) to (x4);
      \draw [dotted,thick] (x1) to (x14);
      \draw [dotted,thick] (x2) to (x24);
      \draw [dotted,thick] (x3) to (x34);
      \draw [dotted,thick] (x12) to (x124);
      \draw [dotted,thick] (x13) to (x134);
      \draw [dotted,thick] (x23) to (x234);
      \draw [dotted,thick] (x123) to (x1234);
   \end{tikzpicture}
   }\\
   \subfloat[Biquadratics pattern of the 3D cube for $\T$]{
      \begin{tikzpicture}
      \node (x) at (0,0) [circle,fill,label=-135:$x$] {};
      \node (x1) at (2,0) [circle,fill,label=-45:$x_{12}$] {};
      \node (x2) at (1,0.75) [circle,draw,ultra thick,label=135:$y_{1}^{\left(2\right)}$] {};
      \node (x12) at (3,0.75) [circle,draw,ultra thick,label=-45:$y_{2}^{\left(2\right)}$] {};
      \node (x3) at (0,2) [circle,draw,ultra thick,label=135:$y_{1}^{\left(1\right)}$] {};
      \node (x13) at (2,2) [circle,draw,ultra thick,label=-45:$y_{2}^{\left(1\right)}$] {};
      \node (x23) at (1,2.75) [circle,fill,label=135:$X$] {};
      \node (x123) at (3,2.75) [circle,fill,label=45:$X_{12}$] {};
      \draw (x) to (x1) to (x12) to (x123) to (x23) to (x3) to (x);
      \draw (x1) to (x13) to (x3);
      \draw (x13) to (x123);
      \draw [dashed] (x) to (x2) to (x12);
      \draw [dashed] (x2) to (x23);
   \end{tikzpicture}
   }\qquad
   \subfloat[Biquadratics pattern of the 3D cube for $\bar{\T}$]{
   \begin{tikzpicture}
      \node (x) at (0,0) [draw,ultra thick,label=-135:$x_{1}$] {};
      \node (x1) at (2,0) [draw,ultra thick,label=-45:$x_{2}$] {};
      \node (x2) at (1,0.75) [fill,label=135:$y^{\left(2\right)}$] {};
      \node (x12) at (3,0.75) [fill,label=-45:$y_{12}^{\left(2\right)}$] {};
      \node (x3) at (0,2) [fill,label=135:$y^{\left(1\right)}$] {};
      \node (x13) at (2,2) [fill,label=-45:$y_{12}^{\left(1\right)}$] {};
      \node (x23) at (1,2.75) [draw,ultra thick,label=135:$X_{1}$] {};
      \node (x123) at (3,2.75) [draw,ultra thick,label=45:$X_{2}$] {};
      \draw (x) to (x1) to (x12) to (x123) to (x23) to (x3) to (x);
      \draw (x1) to (x13) to (x3);
      \draw (x13) to (x123);
      \draw [dashed] (x) to (x2) to (x12);
      \draw [dashed] (x2) to (x23);
   \end{tikzpicture}
   }
   \caption{Two B\"acklund transforms $y^{\left(1\right)}$ and $y^{\left(2\right)}$ of the solution $x$ of a type~\Hvier\ equation in the rhombic version, where both B\"acklund transformations can be described by type~\Hvier\ equations in the trapezoidal version.}
   \label{fig:classi6}
\end{figure}
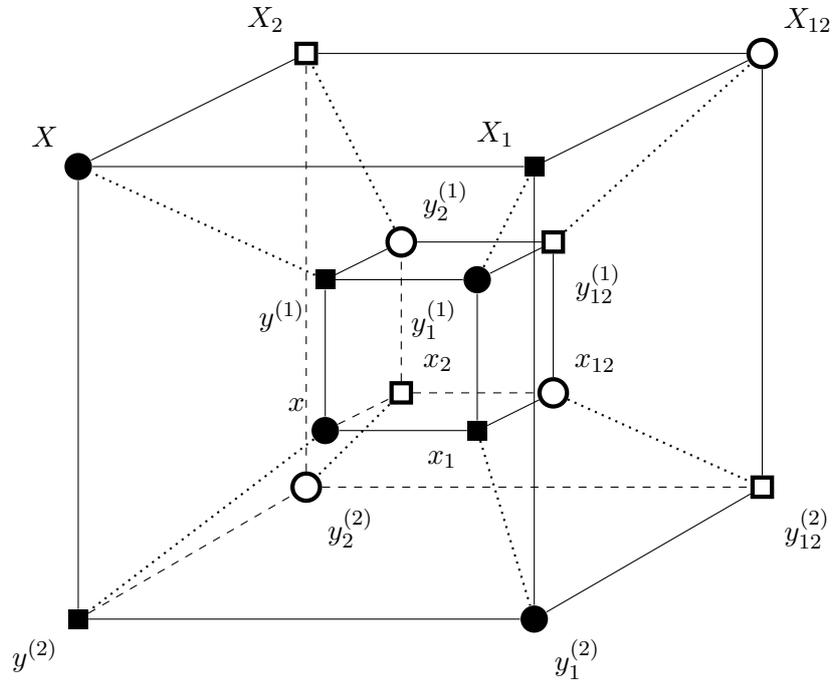
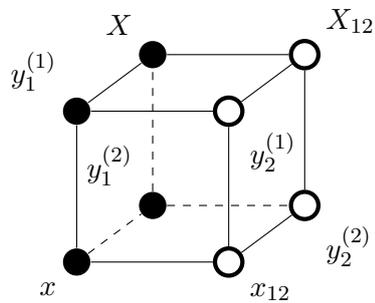
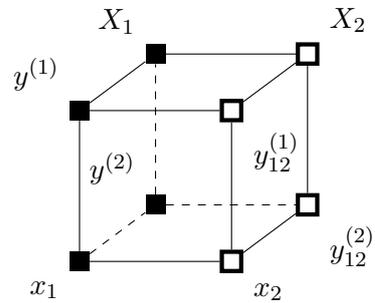
\begin{figure}[htbp]
   \centering
   \subfloat[Biquadratics pattern for the 4D cube]{
   \begin{tikzpicture}
      \node (x) at (0,0) [circle,fill,label=135:$x$] {};
      \node (x1) at (2,0) [fill,label=-135:$x_{1}$] {};
      \node (x2) at (1,0.5) [draw,ultra thick,label=45:$x_{2}$] {};
      \node (x3) at (0,2) [fill,label=-135:$y^{\left(1\right)}$] {};
      \node (x4) at (-3.25,-2.5) [fill,label=-135:$y^{\left(2\right)}$] {};
      \node (x12) at (3,0.5) [circle,draw,ultra thick,label=45:$x_{12}$] {};
      \node (x13) at (2,2) [circle,fill,label=-135:$y_{1}^{\left(1\right)}$] {};
      \node (x14) at (2.75,-2.5) [circle,fill,label=-45:$y_{1}^{\left(2\right)}$] {};
      \node (x23) at (1,2.5) [circle,draw,ultra thick,label=45:$y_{2}^{\left(1\right)}$] {};
      \node (x24) at (-0.25,-0.75) [circle,draw,ultra thick,label=-45:$y_{2}^{\left(2\right)}$] {};
      \node (x34) at (-3.25,3.5) [circle,fill,label=135:$X$] {};
      \node (x123) at (3,2.5) [draw,ultra thick,label=-45:$y_{12}^{\left(1\right)}$] {};
      \node (x124) at (5.75,-0.75) [draw,ultra thick,label=-45:$y_{12}^{\left(2\right)}$] {};
      \node (x134) at (2.75,3.5) [fill,label=135:$X_{1}$] {};
      \node (x234) at (-0.25,5) [draw,ultra thick,label=135:$X_{2}$] {};
      \node (x1234) at (5.75,5) [circle,draw,ultra thick,label=45:$X_{12}$] {};
      \draw (x) to (x1) to (x12) to (x123) to (x23) to (x3) to (x);
      \draw (x4) to (x14) to (x124) to (x1234) to (x234) to (x34) to (x4);
      \draw (x1) to (x13) to (x3);
      \draw (x14) to (x134) to (x34);
      \draw (x13) to (x123);
      \draw (x134) to (x1234);
      \draw [dashed] (x) to (x2) to (x12);
      \draw [dashed] (x4) to (x24) to (x124);
      \draw [dashed] (x2) to (x23);
      \draw [dashed] (x24) to (x234);
      \draw [dotted,thick] (x) to (x4);
      \draw [dotted,thick] (x1) to (x14);
      \draw [dotted,thick] (x2) to (x24);
      \draw [dotted,thick] (x3) to (x34);
      \draw [dotted,thick] (x12) to (x124);
      \draw [dotted,thick] (x13) to (x134);
      \draw [dotted,thick] (x23) to (x234);
      \draw [dotted,thick] (x123) to (x1234);
   \end{tikzpicture}
   }\\
   \subfloat[Biquadratics pattern of the 3D cube for $\T$]{
      \begin{tikzpicture}
      \node (x) at (0,0) [circle,fill,label=-135:$x$] {};
      \node (x1) at (2,0) [circle,draw,ultra thick,label=-45:$x_{12}$] {};
      \node (x2) at (1,0.75) [circle,fill,label=135:$y_{1}^{\left(2\right)}$] {};
      \node (x12) at (3,0.75) [circle,draw,ultra thick,label=-45:$y_{2}^{\left(2\right)}$] {};
      \node (x3) at (0,2) [circle,fill,label=135:$y_{1}^{\left(1\right)}$] {};
      \node (x13) at (2,2) [circle,draw,ultra thick,label=-45:$y_{2}^{\left(1\right)}$] {};
      \node (x23) at (1,2.75) [circle,fill,label=135:$X$] {};
      \node (x123) at (3,2.75) [circle,draw,ultra thick,label=45:$X_{12}$] {};
      \draw (x) to (x1) to (x12) to (x123) to (x23) to (x3) to (x);
      \draw (x1) to (x13) to (x3);
      \draw (x13) to (x123);
      \draw [dashed] (x) to (x2) to (x12);
      \draw [dashed] (x2) to (x23);
   \end{tikzpicture}
   }\qquad
   \subfloat[Biquadratics pattern of the 3D cube for $\bar{\T}$]{
   \begin{tikzpicture}
      \node (x) at (0,0) [fill,label=-135:$x_{1}$] {};
      \node (x1) at (2,0) [draw,ultra thick,label=-45:$x_{2}$] {};
      \node (x2) at (1,0.75) [fill,label=135:$y^{\left(2\right)}$] {};
      \node (x12) at (3,0.75) [draw,ultra thick,label=-45:$y_{12}^{\left(2\right)}$] {};
      \node (x3) at (0,2) [fill,label=135:$y^{\left(1\right)}$] {};
      \node (x13) at (2,2) [draw,ultra thick,label=-45:$y_{12}^{\left(1\right)}$] {};
      \node (x23) at (1,2.75) [fill,label=135:$X_{1}$] {};
      \node (x123) at (3,2.75) [draw,ultra thick,label=45:$X_{2}$] {};
      \draw (x) to (x1) to (x12) to (x123) to (x23) to (x3) to (x);
      \draw (x1) to (x13) to (x3);
      \draw (x13) to (x123);
      \draw [dashed] (x) to (x2) to (x12);
      \draw [dashed] (x2) to (x23);
   \end{tikzpicture}
   }
   \caption{Two B\"acklund transforms $y^{\left(1\right)}$ and $y^{\left(2\right)}$ of the solution $x$ of a type~\Hvier\ equation in the trapezoidal version, where both B\"acklund transformations can be described by type~Q and type~\Hvier\ equations in the trapezoidal version.}
   \label{fig:classi7}
\end{figure}
\begin{figure}[htbp]
   \centering
   \subfloat[Biquadratics pattern for the 4D cube]{
   \begin{tikzpicture}
      \node (x) at (0,0) [circle,fill,label=135:$x$] {};
      \node (x1) at (2,0) [fill,label=-135:$x_{1}$] {};
      \node (x2) at (1,0.5) [draw,ultra thick,label=45:$x_{2}$] {};
      \node (x3) at (0,2) [fill,label=-135:$y^{\left(1\right)}$] {};
      \node (x4) at (-3.25,-2.5) [draw,ultra thick,label=-135:$y^{\left(2\right)}$] {};
      \node (x12) at (3,0.5) [circle,draw,ultra thick,label=45:$x_{12}$] {};
      \node (x13) at (2,2) [circle,fill,label=-135:$y_{1}^{\left(1\right)}$] {};
      \node (x14) at (2.75,-2.5) [circle,draw,ultra thick,label=-45:$y_{1}^{\left(2\right)}$] {};
      \node (x23) at (1,2.5) [circle,draw,ultra thick,label=45:$y_{2}^{\left(1\right)}$] {};
      \node (x24) at (-0.25,-0.75) [circle,fill,label=-45:$y_{2}^{\left(2\right)}$] {};
      \node (x34) at (-3.25,3.5) [circle,draw,ultra thick,label=135:$X$] {};
      \node (x123) at (3,2.5) [draw,ultra thick,label=-45:$y_{12}^{\left(1\right)}$] {};
      \node (x124) at (5.75,-0.75) [fill,label=-45:$y_{12}^{\left(2\right)}$] {};
      \node (x134) at (2.75,3.5) [draw,ultra thick,label=135:$X_{1}$] {};
      \node (x234) at (-0.25,5) [fill,label=135:$X_{2}$] {};
      \node (x1234) at (5.75,5) [circle,fill,label=45:$X_{12}$] {};
      \draw (x) to (x1) to (x12) to (x123) to (x23) to (x3) to (x);
      \draw (x4) to (x14) to (x124) to (x1234) to (x234) to (x34) to (x4);
      \draw (x1) to (x13) to (x3);
      \draw (x14) to (x134) to (x34);
      \draw (x13) to (x123);
      \draw (x134) to (x1234);
      \draw [dashed] (x) to (x2) to (x12);
      \draw [dashed] (x4) to (x24) to (x124);
      \draw [dashed] (x2) to (x23);
      \draw [dashed] (x24) to (x234);
      \draw [dotted,thick] (x) to (x4);
      \draw [dotted,thick] (x1) to (x14);
      \draw [dotted,thick] (x2) to (x24);
      \draw [dotted,thick] (x3) to (x34);
      \draw [dotted,thick] (x12) to (x124);
      \draw [dotted,thick] (x13) to (x134);
      \draw [dotted,thick] (x23) to (x234);
      \draw [dotted,thick] (x123) to (x1234);
   \end{tikzpicture}
   }\\
   \subfloat[Biquadratics pattern of the 3D cube for $\T$]{
      \begin{tikzpicture}
      \node (x) at (0,0) [circle,fill,label=-135:$x$] {};
      \node (x1) at (2,0) [circle,draw,ultra thick,label=-45:$x_{12}$] {};
      \node (x2) at (1,0.75) [circle,draw,ultra thick,label=135:$y_{1}^{\left(2\right)}$] {};
      \node (x12) at (3,0.75) [circle,fill,label=-45:$y_{2}^{\left(2\right)}$] {};
      \node (x3) at (0,2) [circle,fill,label=135:$y_{1}^{\left(1\right)}$] {};
      \node (x13) at (2,2) [circle,draw,ultra thick,label=-45:$y_{2}^{\left(1\right)}$] {};
      \node (x23) at (1,2.75) [circle,draw,ultra thick,label=135:$X$] {};
      \node (x123) at (3,2.75) [circle,fill,label=45:$X_{12}$] {};
      \draw (x) to (x1) to (x12) to (x123) to (x23) to (x3) to (x);
      \draw (x1) to (x13) to (x3);
      \draw (x13) to (x123);
      \draw [dashed] (x) to (x2) to (x12);
      \draw [dashed] (x2) to (x23);
   \end{tikzpicture}
   }\qquad
   \subfloat[Biquadratics pattern of the 3D cube for $\bar{\T}$]{
   \begin{tikzpicture}
      \node (x) at (0,0) [fill,label=-135:$x_{1}$] {};
      \node (x1) at (2,0) [draw,ultra thick,label=-45:$x_{2}$] {};
      \node (x2) at (1,0.75) [draw,ultra thick,label=135:$y^{\left(2\right)}$] {};
      \node (x12) at (3,0.75) [fill,label=-45:$y_{12}^{\left(2\right)}$] {};
      \node (x3) at (0,2) [fill,label=135:$y^{\left(1\right)}$] {};
      \node (x13) at (2,2) [draw,ultra thick,label=-45:$y_{12}^{\left(1\right)}$] {};
      \node (x23) at (1,2.75) [draw,ultra thick,label=135:$X_{1}$] {};
      \node (x123) at (3,2.75) [fill,label=45:$X_{2}$] {};
      \draw (x) to (x1) to (x12) to (x123) to (x23) to (x3) to (x);
      \draw (x1) to (x13) to (x3);
      \draw (x13) to (x123);
      \draw [dashed] (x) to (x2) to (x12);
      \draw [dashed] (x2) to (x23);
   \end{tikzpicture}
   }
   \caption{Two B\"acklund transforms $y^{\left(1\right)}$ and $y^{\left(2\right)}$ of the solution $x$ of a type~\Hvier\ equation in the trapezoidal version, where one B\"acklund transformation can be described by type~Q and type~\Hvier\ equations in the trapezoidal version and the other B\"acklund transformation can be described by type~\Hvier\ equations in rhombic and trapezoidal version.}
   \label{fig:classi8}
\end{figure}
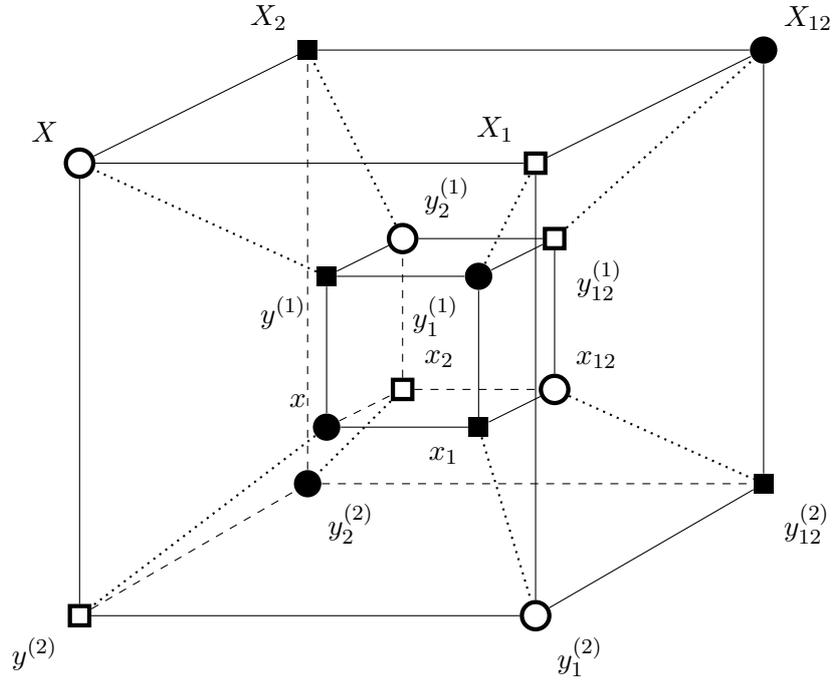
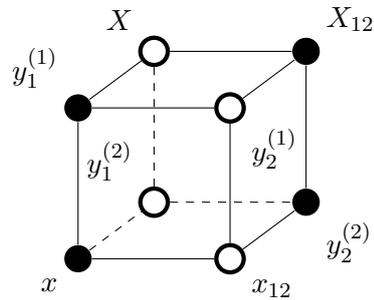
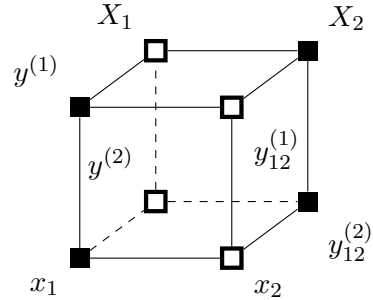
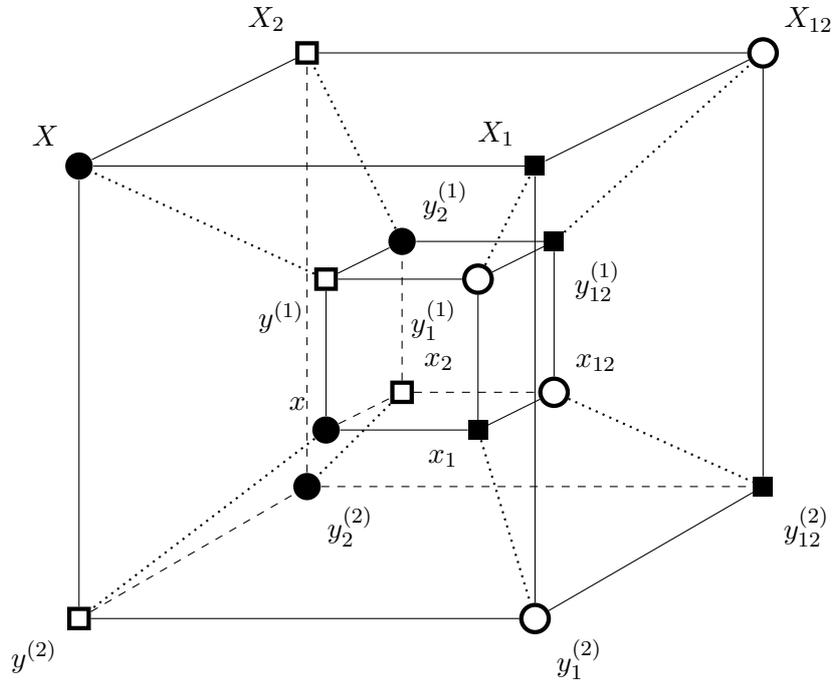
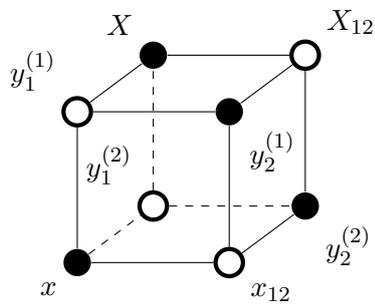
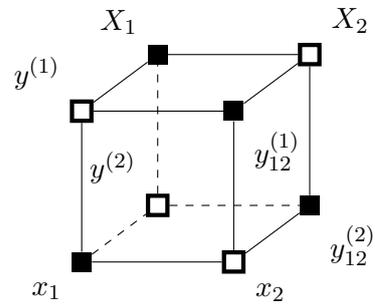
\begin{figure}[htbp]
   \centering
   \subfloat[Biquadratics pattern for the 4D cube]{
   \begin{tikzpicture}
      \node (x) at (0,0) [circle,fill,label=135:$x$] {};
      \node (x1) at (2,0) [fill,label=-135:$x_{1}$] {};
      \node (x2) at (1,0.5) [draw,ultra thick,label=45:$x_{2}$] {};
      \node (x3) at (0,2) [draw,ultra thick,label=-135:$y^{\left(1\right)}$] {};
      \node (x4) at (-3.25,-2.5) [draw,ultra thick,label=-135:$y^{\left(2\right)}$] {};
      \node (x12) at (3,0.5) [circle,draw,ultra thick,label=45:$x_{12}$] {};
      \node (x13) at (2,2) [circle,draw,ultra thick,label=-135:$y_{1}^{\left(1\right)}$] {};
      \node (x14) at (2.75,-2.5) [circle,draw,ultra thick,label=-45:$y_{1}^{\left(2\right)}$] {};
      \node (x23) at (1,2.5) [circle,fill,label=45:$y_{2}^{\left(1\right)}$] {};
      \node (x24) at (-0.25,-0.75) [circle,fill,label=-45:$y_{2}^{\left(2\right)}$] {};
      \node (x34) at (-3.25,3.5) [circle,fill,label=135:$X$] {};
      \node (x123) at (3,2.5) [fill,label=-45:$y_{12}^{\left(1\right)}$] {};
      \node (x124) at (5.75,-0.75) [fill,label=-45:$y_{12}^{\left(2\right)}$] {};
      \node (x134) at (2.75,3.5) [fill,label=135:$X_{1}$] {};
      \node (x234) at (-0.25,5) [draw,ultra thick,label=135:$X_{2}$] {};
      \node (x1234) at (5.75,5) [circle,draw,ultra thick,label=45:$X_{12}$] {};
      \draw (x) to (x1) to (x12) to (x123) to (x23) to (x3) to (x);
      \draw (x4) to (x14) to (x124) to (x1234) to (x234) to (x34) to (x4);
      \draw (x1) to (x13) to (x3);
      \draw (x14) to (x134) to (x34);
      \draw (x13) to (x123);
      \draw (x134) to (x1234);
      \draw [dashed] (x) to (x2) to (x12);
      \draw [dashed] (x4) to (x24) to (x124);
      \draw [dashed] (x2) to (x23);
      \draw [dashed] (x24) to (x234);
      \draw [dotted,thick] (x) to (x4);
      \draw [dotted,thick] (x1) to (x14);
      \draw [dotted,thick] (x2) to (x24);
      \draw [dotted,thick] (x3) to (x34);
      \draw [dotted,thick] (x12) to (x124);
      \draw [dotted,thick] (x13) to (x134);
      \draw [dotted,thick] (x23) to (x234);
      \draw [dotted,thick] (x123) to (x1234);
   \end{tikzpicture}
   }\\
   \subfloat[Biquadratics pattern of the 3D cube for $\T$]{
      \begin{tikzpicture}
      \node (x) at (0,0) [circle,fill,label=-135:$x$] {};
      \node (x1) at (2,0) [circle,draw,ultra thick,label=-45:$x_{12}$] {};
      \node (x2) at (1,0.75) [circle,draw,ultra thick,label=135:$y_{1}^{\left(2\right)}$] {};
      \node (x12) at (3,0.75) [circle,fill,label=-45:$y_{2}^{\left(2\right)}$] {};
      \node (x3) at (0,2) [circle,draw,ultra thick,label=135:$y_{1}^{\left(1\right)}$] {};
      \node (x13) at (2,2) [circle,fill,label=-45:$y_{2}^{\left(1\right)}$] {};
      \node (x23) at (1,2.75) [circle,fill,label=135:$X$] {};
      \node (x123) at (3,2.75) [circle,draw,ultra thick,label=45:$X_{12}$] {};
      \draw (x) to (x1) to (x12) to (x123) to (x23) to (x3) to (x);
      \draw (x1) to (x13) to (x3);
      \draw (x13) to (x123);
      \draw [dashed] (x) to (x2) to (x12);
      \draw [dashed] (x2) to (x23);
   \end{tikzpicture}
   }\qquad
   \subfloat[Biquadratics pattern of the 3D cube for $\bar{\T}$]{
   \begin{tikzpicture}
      \node (x) at (0,0) [fill,label=-135:$x_{1}$] {};
      \node (x1) at (2,0) [draw,ultra thick,label=-45:$x_{2}$] {};
      \node (x2) at (1,0.75) [draw,ultra thick,label=135:$y^{\left(2\right)}$] {};
      \node (x12) at (3,0.75) [fill,label=-45:$y_{12}^{\left(2\right)}$] {};
      \node (x3) at (0,2) [draw,ultra thick,label=135:$y^{\left(1\right)}$] {};
      \node (x13) at (2,2) [fill,label=-45:$y_{12}^{\left(1\right)}$] {};
      \node (x23) at (1,2.75) [fill,label=135:$X_{1}$] {};
      \node (x123) at (3,2.75) [draw,ultra thick,label=45:$X_{2}$] {};
      \draw (x) to (x1) to (x12) to (x123) to (x23) to (x3) to (x);
      \draw (x1) to (x13) to (x3);
      \draw (x13) to (x123);
      \draw [dashed] (x) to (x2) to (x12);
      \draw [dashed] (x2) to (x23);
   \end{tikzpicture}
   }
   \caption{Two B\"acklund transforms $y^{\left(1\right)}$ and $y^{\left(2\right)}$ of the solution $x$ of a type~\Hvier\ equation in the trapezoidal version, where both B\"acklund transformations can be described by type~\Hvier\ equations in rhombic and trapezoidal version.}
   \label{fig:classi9}
\end{figure}
\newpage

\bibliographystyle{amsalpha}
\bibliography{Quellen}

\end{document}